\documentclass[12pt]{article}

\usepackage[utf8]{inputenc}
\usepackage{alphabeta}
\usepackage{amsmath}
\usepackage{amssymb}
\usepackage{mathrsfs}
\usepackage{graphicx}
\usepackage{bbold}
\usepackage{comment}
\usepackage{transparent}
\usepackage{bold-extra}
\usepackage[dvipsnames]{xcolor}
\usepackage[most]{tcolorbox}
\usepackage{textcomp}

\usepackage{braket}
\usepackage{bbm}
\usepackage{physics}
\usepackage{comment}
\usepackage{listings}

\numberwithin{equation}{section}

\renewcommand{\lstlistingname}{\bfseries Listing}
\makeatletter
\def\fnum@lstlisting{%
  \lstlistingname
  \ifx\lst@@caption\@empty\else~\thelstlisting\normalfont\fi}%
\makeatother

\definecolor{codegreen}{rgb}{0,0.6,0}
\definecolor{codegray}{rgb}{0.5,0.5,0.5}
\definecolor{codepurple}{rgb}{0.58,0,0.82}
\definecolor{backcolour}{rgb}{0.95,0.95,0.92}
\definecolor{doccolour}{rgb}{0.92,0.94,0.96}

\lstdefinestyle{mystyle}{
    backgroundcolor=\color{backcolour},   
    commentstyle=\color{codegreen},
    keywordstyle=\color{magenta},
    numberstyle=\tiny\color{codegray},
    stringstyle=\color{codepurple},
    basicstyle=\ttfamily\footnotesize,
    breakatwhitespace=false,         
    breaklines=true,                 
    captionpos=b,                    
    keepspaces=true,                 
    numbers=left,                    
    numbersep=5pt,                  
    showspaces=false,                
    showstringspaces=false,
    showtabs=false,                  
    tabsize=2
}

\usepackage{braket}
\usepackage{bm}
\usepackage{bbm}
\usepackage{hyperref}
\numberwithin{equation}{section}

\newcommand{\mycomment}[1]{}

\newcommand{\fig}[1]{Figure~\ref{#1}}

\usepackage{subfig}
\makeatother

\begin{document}
\begin{titlepage}
\renewcommand{\thefootnote}{\fnsymbol{footnote}}

\begin{flushright} 
  
\end{flushright} 

\vspace{1.5cm}

\begin{center}
  {\bf \large
  Tensor renormalization group study of the three-dimensional SU(2) and SU(3) gauge theories with the reduced tensor network formulation
  }
\end{center}

\vspace{1cm}


\begin{center}
         Atis Y{\sc osprakob}$^{1)}$\footnote
          { E-mail address : ayosp(at)phys.sc.niigata-u.ac.jp}
          and
          Kouichi O{\sc kunishi}$^{1)}$\footnote
          { E-mail address : okunishi(at)phys.sc.niigata-u.ac.jp}


\vspace{1cm}

$^{1)}$\textit{Department of Physics, Niigata University, Niigata 950-2181, Japan}

\end{center}

\vspace{0.5cm}

\begin{abstract}
  \noindent

We perform a tensor renormalization group simulation of non-Abelian gauge theory in three dimensions using a formulation based on the `armillary sphere.' In this formulation, matrix indices are completely traced out, eliminating the degeneracy in the singular value spectrum of the initial tensor. We demonstrate the usefulness of this technique by computing the average plaquette at zero temperature and the Polyakov loop susceptibility at finite temperatures for 2+1D SU(2) and SU(3) gauge theories. The deconfinement transition is identified for both gauge groups, with the SU(2) case being consistent with previous Monte Carlo results.

\end{abstract}
\vfill
\end{titlepage}
\vfil\eject


\renewcommand{\thefootnote}{\arabic{footnote}}
\setcounter{footnote}{0}


\section{Introduction}

Lattice quantum chromodynamics (QCD) is of particular interest theoretically and numerically due to its rich mathematical structure and challenging technical setups. As a four-dimensional non-Abelian gauge theory with multiple flavors of fermions, QCD exhibits many phases on its extensive parameter space. Many of these phases such as the finite-density QCD and nonzero topological angles are inaccessible to the Monte Carlo methods due to the sign problem \cite{Nagata:2021ugx}. Although some techniques; e.g., the complex Langevin method \cite{Parisi:1983mgm, Aarts:2009uq, Scherzer:2018hid, Nagata:2015uga} or the Lefschetz thimble method \cite{Picard1897,Lefschetz1924,Witten:2010cx, Witten:2010zr,Fukuma:2020fez, Fujisawa:2021hxh}, have been developed to address this issue and have indeed shown promising results, these methods still have a lot of practical limitations.

In recent years, the tensor normalization group (TRG) approach has been considered as an alternative to the Monte Carlo method because it is free from the sign problem by construction. Furthermore, because it utilizes a coarse-graining procedure, its computational costs only grow logarithmically with volumes \cite{Levin:2006jai,PhysRevLett.115.180405,Adachi:2020upk}. The Grassmann variation of the method can also treat fermionic variables directly without the need to integrate them first \cite{Gu:2013gba,Sakai:2017jwp,Akiyama:2020sfo,Yosprakob:2023flr}. Although the TRG method was first developed for two-dimensional systems, higher-dimensional versions were later proposed and shown to apply to three- and four-dimensional systems \cite{PhysRevB.86.045139,Adachi:2019paf,Kadoh:2019kqk}.

Although simulating a full lattice QCD with the TRG method is still too challenging, there has been some progress for pure U($N$) and SU($N$) gauge theories \cite{Fukuma:2021cni,Hirasawa:2021qvh,Kuwahara:2022ubg} (see also \cite{Akiyama:2024qgv} for 3D SU(2) principal chiral model). In these studies, the methods for constructing the tensor network for non-Abelian gauge theory fall into two categories; the sampling-based method and the character expansion. In the sampling-based method, the Haar measure for the group integral is approximated by a Monte Carlo sampling \cite{Fukuma:2021cni,Kuwahara:2022ubg}. This method avoids the discretization of the group manifold, making it a convenient option for complicated actions. However, the precision of the method strongly depends on the ensemble of the samples, which can be unknown in some parameter regions. On the other hand, one can construct the tensor network in the dual formalism with the character expansion \cite{Liu:2013nsa,Hirasawa:2021qvh,Akiyama:2024qgv}. This method does not depend on any meta-parameter,
but the larger number of expansion terms is typically required for smaller couplings.
The two methods were shown to be equivalent---providing the same singular value spectrum in the infinite-$D_\text{cut}$ limit \cite{Fukuma:2021cni}. The spectrum, however, has a severe degeneracy due to the internal symmetry. In the character expansion approach, this degeneracy can be eliminated in two dimensions because the `loops' associated with the internal symmetry are manifested and can be traced out \cite{Hirasawa:2021qvh}. This was later shown to be true also in higher dimensions with the `armillary sphere' formalism \cite{Yosprakob:2023jgl}.

The goal of this paper is to demonstrate the usefulness of this formalism in the three-dimensional SU(2) and SU(3) gauge theories which are the simplest non-trivial examples. Here, the average plaquette at zero temperature is computed and compared with the strong coupling prediction, with a satisfactory agreement. We also identify the deconfinement at finite temperatures for both gauge groups by computing the Polyakov loop susceptibility.

This paper is organized as follows. In section \ref{section:formulation}, we give a brief review of the three-dimensional SU($N$) gauge theory and the construction of the armillary tensor. In section \ref{section:results}, we test our method for both SU(2) and SU(3) at zero and finite temperature. Section \ref{section:summary} is devoted to summary and discussion. The definition of the 3-legged Clebsch-Gordan coefficient is given in appendix \ref{sec:gCGs}. The list of multi-representation indices for SU(2) and SU(3) gauge groups are given in appendix \ref{sec:mri}.

\section{Formulation}
\label{section:formulation}
\subsection{Character expansion for non-Abelian gauge theory}

The three-dimensional SU($N$) gauge theory is defined by the action
\begin{equation}
    S = \frac{1}{2g^2}\int d^3x \tr(F_{\mu\nu}F^{\mu\nu}) \label{eq:continuum_action}
\end{equation}
where $g$ is the gauge coupling constant and $F_{\mu\nu}$ ($\mu,\nu=1,2,3$) is the field strength tensor
\begin{equation}
    F_{\mu\nu}=\partial_\mu A_\nu-\partial_\nu A_\mu-i[A_\mu,A_\nu].
\end{equation}
The gauge field $A_\mu$ belongs to the fundamental representation of the algebra $\mathfrak{su}(N)$.

To define this theory on a lattice, we consider the three-dimensional periodic lattice $\Lambda_3$ where the Euclidean space-time is discretized with $x_\mu=an_\mu$ with $n\in\Lambda_3$ and lattice spacing $a$. We define a link variable $U_\mu(n)\in\text{SU}(N)$ on a link between the sites $n$ and $n+\hat\mu$, which corresponds to the gauge field as $U_\mu(n)\sim\exp(iaA_\mu(x))$. In this way, one can define the plaquette variable
\begin{equation}
    P_{\mu\nu}(n)=U_{\mu}(n)U_{\nu}(n+\hat\mu)U^\dagger_{\mu}(n+\hat\nu) U^\dagger_{\nu}(n)
    \label{eq:plaquette}
\end{equation}
which corresponds to the field strength tensor via $P_{\mu\nu}(n)\sim \exp(ia^2F_{\mu\nu}(x))$. The simplest gauge-invariant regularization of \eqref{eq:continuum_action} on the lattice is given by
\begin{equation}
    S = S_P =\frac{\beta}{N}\sum_{n\in\Lambda_3}\sum_{\mu\nu\in\{13,21,32\}}\Re\tr(\mathbb{1}-P_{\mu\nu}(n)).
\end{equation}
The parameter $\beta$ is related to the gauge coupling $g$ via $\beta=2N/(ag^2)$. 

The corresponding partition function is
\begin{equation}
    Z=\int dU e^{-S} = \int dU\prod_{n\in\Lambda_3}\prod_{\mu\nu\in\{13,21,32\}}e^{\frac{\beta}{N}\Re\tr(P_{\mu\nu}(n)-\mathbb{1})}.
\end{equation}

The construction of the tensor network with character expansions can be done by performing the expansion on the local Boltzmann weight
\begin{equation}
    e^{\frac{\beta}{N}\Re\tr(P-\mathbb{1})}=\sum_r f_r(\beta) \tr_r P
    \label{eq:character_expansion}
\end{equation}
where the sum is taken over all irreducible representations (which we will simply refer to as `representations' from now on) of SU($N$) and the symbol $\tr_r$ represents the trace over the group element $P$ under the representation $r$; i.e., the group character. We can label a representation $r$ of SU($N$) by a tuple of $N$ non-negative integers
\begin{equation}
    r = (l_1,l_2,\cdots,l_{N})
    \label{eq:irrep_labeling}
\end{equation}
satisfying
\begin{equation}
    l_1\geq l_2\geq \cdots\geq l_{N-1}\geq l_N\equiv 0.
\end{equation}
This is equivalently represented by the Young tableaux with $l_i$ boxes in the $i$-th row. In terms of this labeling, the character expansion coefficient can be evaluated by \cite{Drouffe:1983fv}
\begin{equation}
    f_r(\beta)=\int dP e^{\frac{\beta}{N}\Re\tr(P-\mathbb{1})}\tr_r P^\dagger =e^{-\beta}\sum_{q\in\mathbb{Z}}\det\left(I_{l_j+i-j+q}(\beta/N)\right)_{ij}
    \label{eq:expansion_coefficient}
\end{equation}
where $I_n(z)$ is the modified Bessel function and the dimensionality of the representation $r$ is given by
\begin{equation}
    d_r=\prod_{1\leq i<j\leq N}\left(1+\frac{l_i-l_j}{j-i}\right).
\end{equation}

The plaquette trace in \eqref{eq:character_expansion} can be decomposed into the product of link variables
\begin{equation}
    \tr_rP_{\mu\nu}(n)=\sum_{i,j,k,l}
    (U_{\mu}(n))^r_{ij}
    (U_{\nu}(n+\hat\mu))^r_{jk}
    (U^\dagger_{\mu}(n+\hat\nu))^r_{kl}
    (U^\dagger_{\nu}(n))^r_{li},
\end{equation}
where $U^r_{ij}$ is the matrix element of $U$ under the representation $r$. We can now perform the group integral of each link variable separately (indices on the left-hand side are suppressed)
\begin{equation}
    z_{\mu}(n)=\int dU_{\mu}(n) (U_{\mu}(n))^{r_1}_{i_1i'_1}(U_{\mu}(n))^{r_2}_{i_2i'_2}(U^\dagger_{\mu}(n))^{s_1}_{j_1j'_1}(U^\dagger_{\mu}(n))^{s_2}_{j_2j'_2}
    \label{eq:group_integral}
\end{equation}
so that the partition function is now given by
\begin{equation}
    Z = \sum_{\{r\}}\Tr\left\{\left(\prod_{n\in\Lambda_3}\prod_{\mu=1}^3z_\mu(n)\right)\left(\prod_{n\in\Lambda_3}f_{r_n}\right)\right\}.
\end{equation}
Here, the symbol $\Tr$ denotes the appropriate contraction of all the matrix indices.

\subsection{Group integral and the armillary sphere}

To evaluate the group integral \eqref{eq:group_integral}, we rewrite the integrand in terms of irreducible representations using the Clebsch-Gordan decomposition. Namely, consider the products
\begin{align}
    r_1\otimes r_2&=\rho_1\oplus\rho_2\oplus\cdots\oplus\rho_m,\\
    s_1\otimes s_2&=\sigma_1\oplus\sigma_2\oplus\cdots\oplus\sigma_n.
\end{align}
The matrix form of such decomposition is
\begin{align}
    (U)^{r_1}_{i_1i'_1}(U)^{r_2}_{i_2i'_2}&=\sum_{\rho\in \{\rho_1,\cdots,\rho_m\}}\sum_{i,i'}C^{\rho i}_{r_1i_1;r_2i_2}C^{\rho i'}_{r_1i'_1;r_2i'_2}(U)^{\rho}_{ii'},\label{eq:cgdecomposition}\\
    (U^\dagger)^{s_1}_{j_1j'_1}(U^\dagger)^{s_2}_{j_2j'_2}&=\sum_{\sigma\in \{\sigma_1,\cdots,\sigma_n\}}\sum_{j,j'}C^{\sigma j}_{s_1j_1;s_2j_2}C^{\sigma j'}_{s_1j'_1;s_2j'_2}(U^\dagger)^{\sigma}_{jj'}.
\end{align}
A C++ program\footnote{See also \url{https://homepages.physik.uni-muenchen.de/~vondelft/Papers/ClebschGordan/}} provided by Ref. \cite{Alex:2010wi} is used to compute the SU(2) and SU(3) CG coefficients in this work.

\begin{figure}
    \centering
    \includegraphics[scale=0.8]{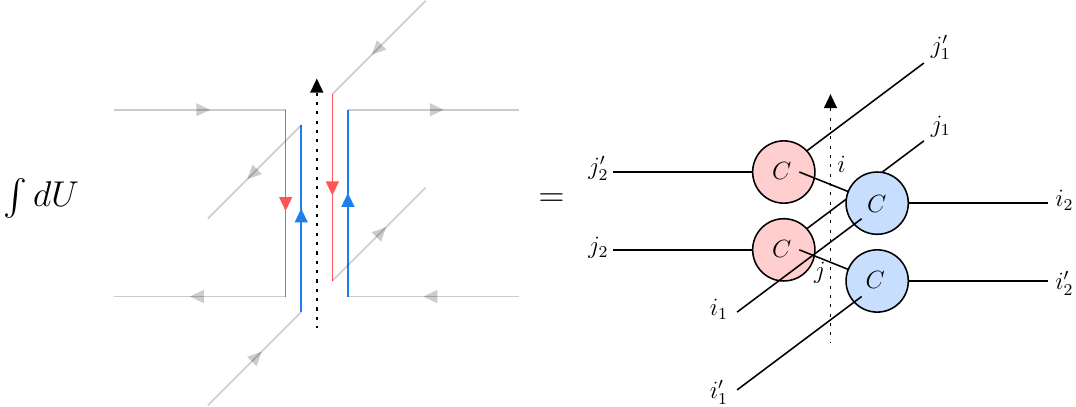}
    \caption{The diagrammatic representation of the group integral \eqref{eq:group_integral_evaluated}. The dashed arrow indicates the orientation of the link variable. Note how the CG coefficients form a structure with two separated layers of matrix indices (although they are still connected via the representations). The CG coefficient tensors are color-coded based on whether they come from the decomposition of $U$ (blue) or $U^\dagger$ (red).}
    \label{fig:group_integral}
\end{figure}

Using this result together with the Schur orthogonality relation
\begin{equation}
    \int dU (U)^r_{ij}(U)^s_{kl}=\frac{1}{d_r}\delta(r\simeq s)\delta_{il}\delta_{jk}
\end{equation}
the group integral \eqref{eq:group_integral} can be evaluated to \cite{Yosprakob:2023jgl,Oeckl:2000hs}
\begin{equation}
    z = \sum_{\rho\in \{\rho_1,\cdots,\rho_m\}}\sum_{\sigma\in \{\sigma_1,\cdots,\sigma_n\}}\frac{1}{d_\rho}\delta(\rho\simeq\sigma)
    \left(\sum_{i}
    C^{\rho i}_{r_1i_1;r_2i_2}C^{\sigma i}_{s_1j'_1;s_2j'_2}\right)\left(\sum_jC^{\rho j}_{r_1i'_1;r_2i'_2}
    C^{\sigma j}_{s_1j_1;s_2j_2}\right).
    \label{eq:group_integral_evaluated}
\end{equation}
The diagrammatic representation of \eqref{eq:group_integral_evaluated} is shown in \fig{fig:group_integral}.
Note that $\delta(\rho\simeq\sigma)$ requires $\rho$ and $\sigma$ to be \emph{isomorphic}. This is a weaker version of $\delta_{\rho,\sigma}$ where $\rho$ and $\sigma$ are required to belong to the \emph{same} vector space. The main difference is that $\delta(\rho\simeq\sigma)$ allows all multiplicities of $\rho$ and $\sigma$ while $\delta_{\rho,\sigma}$ also requires the multiplicity index of $\rho$ and $\sigma$ to match.

\begin{figure}
    \centering
    \includegraphics[scale=0.7]{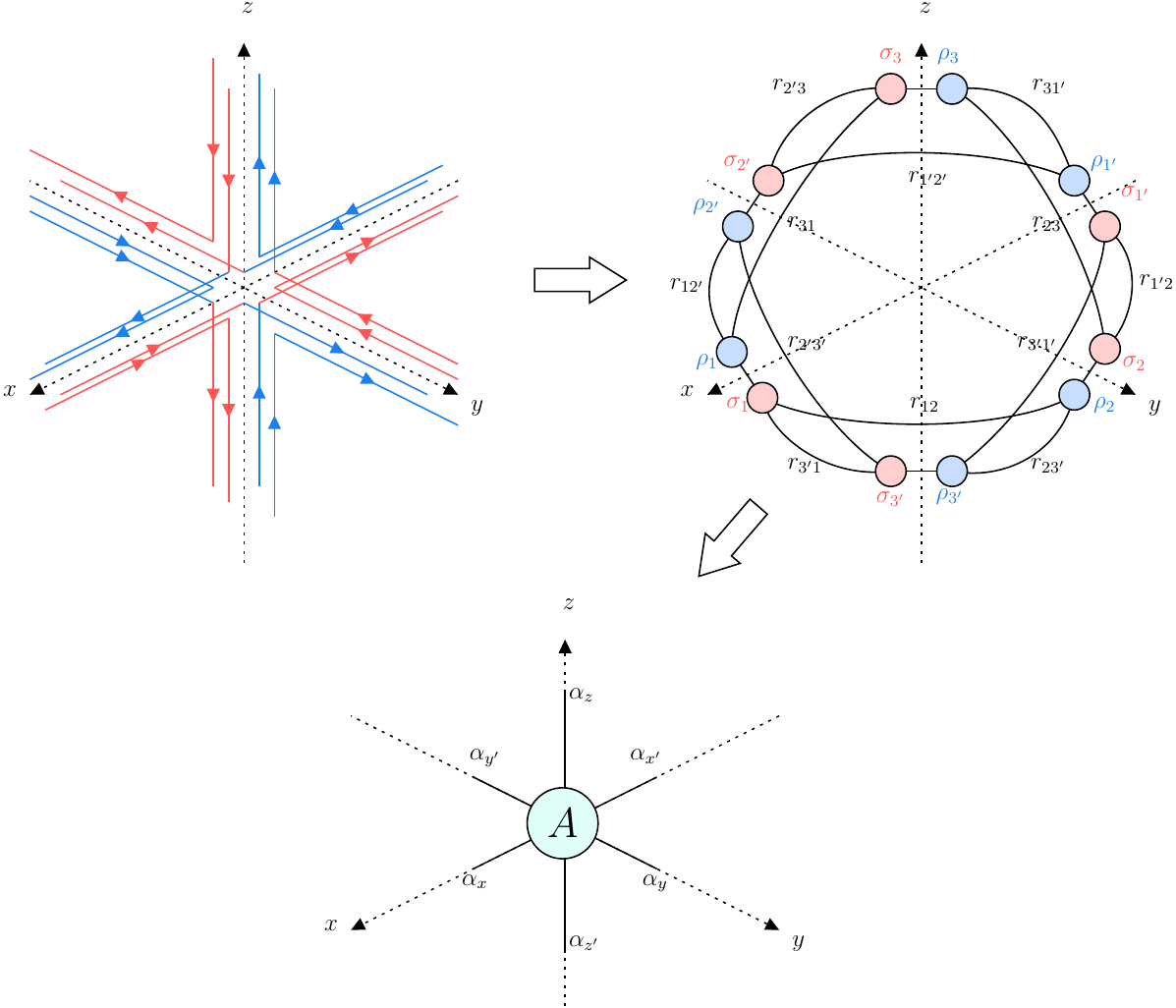}
    \caption{(Top-left) Link variables around the site $n$ before the integration.
    (Top-right) The armillary sphere with the representation associated with each leg shown. Note that the legs connecting the CG coefficients in this diagram represent the contraction of matrix indices only---representation indices are not summed at this step. (Bottom) The armillary tensor \eqref{eq:pure_armillary} with its legs defined in \eqref{eq:mri1}-\eqref{eq:mri2}.
    }
    \label{fig:indices}
\end{figure}

In the reduced tensor network formulation, the CG coefficients form a closed tensor network around each site, which we call the armillary sphere \cite{Yosprakob:2023jgl}, where the matrix indices are completely contracted, as shown in figure \ref{fig:indices}.
An armillary sphere centered at the site $n$ is given by
\begin{align}
    A_{\alpha_x\alpha_{x'}\alpha_y\alpha_{y'}\alpha_z\alpha_{z'}}=\Tr\{\;&   C^{\rho_1}_{r_{13};r_{12'}}C^{\sigma_1}_{r_{13'};r_{12}}
                C^{\rho_2}_{r_{12};r_{23'}}C^{\sigma_2}_{r_{1'2};r_{23}}
                C^{\rho_3}_{r_{23};r_{1'3}}C^{\sigma_3}_{r_{2'3};r_{13}}\nonumber\\
            &   C^{\rho_{1'}}_{r_{1'2'};r_{1'3}}C^{\sigma_{1'}}_{r_{1'2};r_{1'3'}}
                C^{\rho_{2'}}_{r_{2'3'};r_{12'}}C^{\sigma_{2'}}_{r_{2'3};r_{1'2'}}
                C^{\rho_{3'}}_{r_{1'3'};r_{23'}}C^{\sigma_{3'}}_{r_{13'};r_{2'3'}}\,\}
                \label{eq:pure_armillary}\\
            &   \times\prod_{\mu=1,2,3}\delta(\rho_{\mu}\simeq\sigma_{\mu})\delta(\rho_{\mu'}\simeq\sigma_{\mu'})
            \nonumber
\end{align}
where the symbols $r_{\mu\nu}$, $r_{\mu'\nu}$, $r_{\mu\nu'}$, and $r_{\mu'\nu'}$ stand for the representations associated with the plaquettes $P_{\mu\nu}(n)$, $P_{\mu\nu}(n-\hat\mu)$, $P_{\mu\nu}(n-\hat\nu)$, and $P_{\mu\nu}(n-\hat\mu-\hat\nu)$, respectively. The symbol $C^\rho_{r_1r_2}$ is the shorthand notation for the CG coefficient $C^{\rho j}_{r_1i_1;r_2i_2}$. The trace is taken over the matrix indices of the CG coefficients while representation indices are left as free indices. We also define the multi-representation indices\footnote{
The irreps $\rho$ and $\sigma$ in the multi-representation indices should not be viewed as a representation by itself but as an irreducible component of the CG decomposition. This distinction is important when the decomposition gives several copies of equivalent irreps. Then we have to specify not only which irrep it is, but also its multiplicity index. For example, see entries 47 and 48 of Table \ref{tab:su2_poly}.
}
\begin{align}
    \alpha_x &= (r_{31},r_{12'};r_{3'1},r_{12};\rho_1,\sigma_1),\label{eq:mri1}\\
    \alpha_y &= (r_{12},r_{23'};r_{1'2},r_{23};\rho_2,\sigma_2),\\
    \alpha_z &= (r_{23},r_{31'};r_{2'3},r_{31};\rho_3,\sigma_3),\\
    \alpha_{x'} &= (r_{31'},r_{1'2'};r_{3'1'},r_{1'2};\rho_{1'},\sigma_{1'}),\\
    \alpha_{y'} &= (r_{12'},r_{2'3'};r_{1'2'},r_{2'3};\rho_{2'},\sigma_{2'}),\\
    \alpha_{z'} &= (r_{23'},r_{3'1'};r_{2'3'},r_{3'1};\rho_{3'},\sigma_{3'}),\label{eq:mri2}
\end{align}
so that $A$ is a tensor with $\alpha_x,\cdots,\alpha_{z'}$ as its indices. Note that $\rho_\mu$ and $\sigma_\mu$ are required to be isomorphic due to the Kronecker delta in \eqref{eq:group_integral_evaluated}. The constraint given by the Kronecker delta makes the effective bond dimensions for the multi-representation indices small. For example, the multi-representation indices of the $\text{SU}(2)$ gauge theory with 3 expansion terms $\{\textbf{1},\textbf{2},\textbf{3}\}$ is of dimension 46, while the $\text{SU}(3)$ with 3 terms $\{\textbf{1},\textbf{3},\bar{\textbf{3}}\}$ is of dimension 33. The lists of all representation indices for both gauge groups are given in appendix \ref{sec:mri}.

The armillary tensor $A_{\alpha_x\alpha_{x'}\alpha_y\alpha_{y'}\alpha_z\alpha_{z'}}$ is completely group-theoretical. It, therefore, needs to be evaluated only once at the beginning. To obtain the site tensor, one only needs to multiply the armillary tensor with the character expansion coefficients and the dimensional factors from the group integral \eqref{eq:group_integral_evaluated}:
\begin{equation}
    T_{\alpha_x\alpha_{x'}\alpha_y\alpha_{y'}\alpha_z\alpha_{z'}}=A_{\alpha_x\alpha_{x'}\alpha_y\alpha_{y'}\alpha_z\alpha_{z'}}
    \times\frac{(f_{r_{12}}(\beta)f_{r_{1'2}}(\beta)\cdots f_{r_{3'1'}}(\beta))^{1/4}}{\sqrt{
    d_{\rho_{1}}d_{\rho_{2}}d_{\rho_{3}}
    d_{\rho_{1'}}d_{\rho_{2'}}d_{\rho_{3'}}
    }}
    \label{eq:initial;_tensor}
\end{equation}
The plaquette contains 4 sites, thus the 4-th root of the expansion coefficient. And similarly, each link contains 2 sites, thus the square root of the dimensional factor.

\section{Numerical results}
\label{section:results}

\subsection{Average plaquette}

\begin{figure}
    \centering
    \includegraphics[scale=0.8]{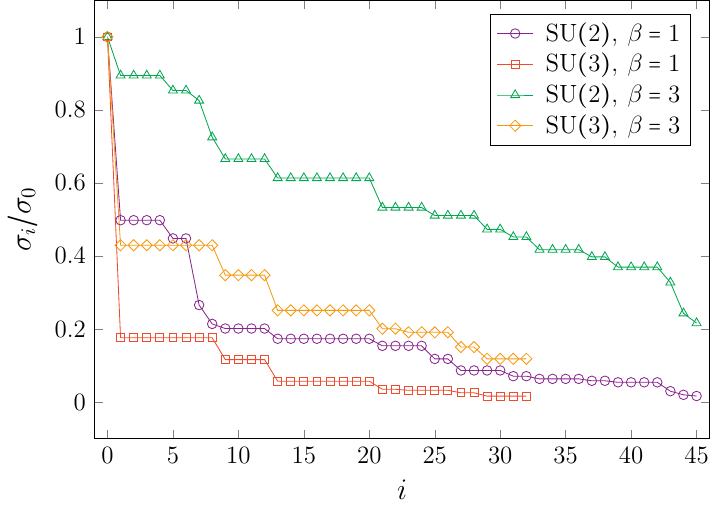}
    \caption{The full singular value spectrum of the initial tensors \eqref{eq:initial;_tensor} at $\beta=1$ and $\beta=3$. The spectrum is obtained from performing the HOSVD on one of the 6 legs, which are all identical due to the symmetry of the tensor.}
    \label{fig:spectrum}
\end{figure}

In this work, we mainly consider the pure SU(2) and SU(3) gauge theories, although the procedure can be straightforwardly generalized for other gauge groups. We follow the guideline given in Ref. \cite{Hirasawa:2021qvh} for choosing the representations in the character expansion; i.e., prioritizing representations with lower dimensionality and always keeping both the representation and its conjugate.
As such, we consider at most 3 expansion terms corresponding to the representations $\mathbf{1}=(0)$, $\mathbf{2}=(1)$, and $\mathbf{3}=(2)$ for SU(2) and $\mathbf{1}=(0,0)$, $\mathbf{3}=(1,0)$, and $\bar{\mathbf{3}}=(1,1)$ for SU(3). As shown in figure \ref{fig:spectrum}, we do not observe severe degeneracy in the spectrum like in Ref. \cite{Fukuma:2021cni} since the internal indices are already traced out.

\begin{figure}
    \centering
    \includegraphics[scale=0.8]{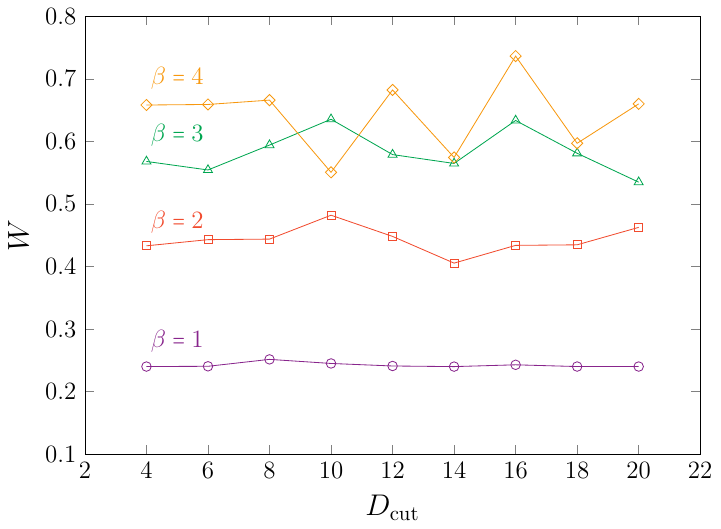}
    \caption{The $D_\text{cut}$ dependence of the average plaquette expectation value of the $\text{SU}(2)$ gauge theory up to $D_\text{cut}=20$.}
    \label{fig:wilson_dcut}
\end{figure}

After obtaining the armillary tensor, we perform the truncation on the site tensor based on the HOSVD \cite{doi:10.1137/S0895479896305696,Xie:2012mjn,Kuwahara:2022ubg} with the truncated bond dimension of $\chi$. The coarse-graining procedure is done with the anisotropic TRG method \cite{Adachi:2019paf} with bond dimension $D_\text{ATRG}$ recursively along the $x$-$y$-$z$ axes up to $V=16^3=4096$. Figure \ref{fig:wilson_dcut} shows the bond-dimension dependence of the average plaquette
\begin{align}
    W &= \frac{1}{N}\langle\tr P\rangle = 1+\frac{1}{3V}\frac{\partial}{\partial\beta}\log Z,\label{eq:average_plaquette}\\
    \frac{\partial}{\partial\beta}\log Z &\approx \left.\frac{1}{2\epsilon}(\log Z(\beta+\epsilon)-\log Z(\beta-\epsilon))\right|_{\epsilon=0.1}.
\end{align}
of the SU(2) gauge theory with $\beta=1,2,3,4$. Here, we show the result with bond dimensions $\chi=D_\text{ATRG}=D_\text{cut}$ up to $D_\text{cut}=20$. The expectation value for $\beta=1$ already converges at $D_\text{cut}=12$. At larger $\beta$, the expectation values fluctuate around a certain value after $D_\text{cut}=14$. As such, we pick $D_\text{cut}=16$ for the rest of the calculations involving the average plaquette.

\begin{figure}
    \centering
    \includegraphics[scale=0.65]{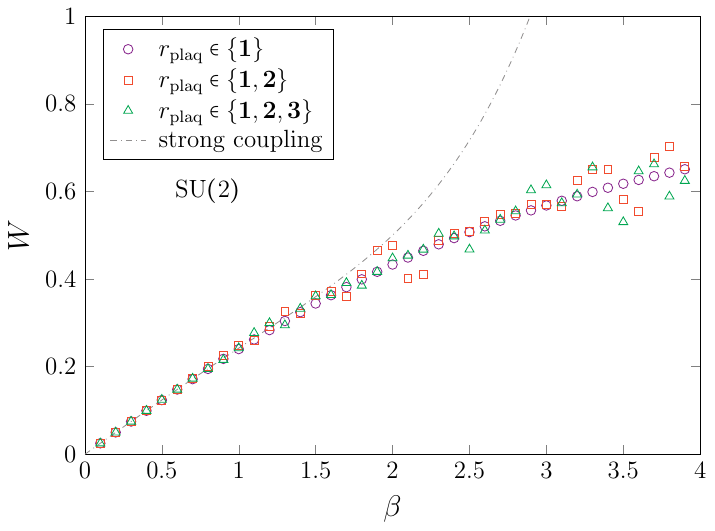}
    \includegraphics[scale=0.65]{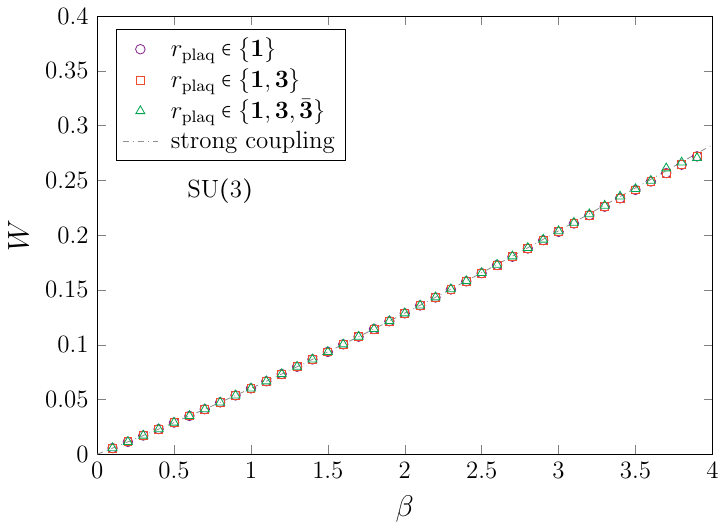}
    \caption{The average plaquette \eqref{eq:average_plaquette} for SU(2) (left) and SU(3) (right) gauge theories as a function of inverse coupling $\beta$. Different markers indicate different numbers of terms in the character expansion. The dashed line is the strong-coupling approximation \eqref{eq:strong_coupling_approximation}.}
    \label{fig:wilson}
\end{figure}

Next, we performed the computation of the average plaquette for both $\text{SU}(2)$ and $\text{SU}(3)$ gauge groups up to $\beta=4$ and compared the results with the strong-coupling approximations in figure \ref{fig:wilson}. The strong-coupling approximation of the average plaquette is given by \cite{Balian:1974xw,Karliner:1987cu}
\begin{equation}
    W=\left\{
    \begin{array}{ll}
    \displaystyle
    \frac{\beta}{4}-\frac{\beta^3}{96}+\frac{\beta^5}{384}+\text{O}(\beta^7)&; \text{SU}(2),\\[3mm]
    \displaystyle
    \frac{\beta}{18}+\frac{\beta^2}{216}-\frac{5\beta^4}{93312}+\text{O}(\beta^5)&; \text{SU}(3).
    \end{array}
    \right.
    \label{eq:strong_coupling_approximation}
\end{equation}
We also did the calculation with various numbers of terms in the character expansion. We find that the average plaquette can be very well approximated by only one expansion term corresponding to the trivial representation, $\mathbf{1}$, in the strong coupling regime for both gauge groups.

In figure \ref{fig:wilson}, when only the trivial representation is included (purple circle), the data does not fluctuate regardless of the coupling. This is because the initial tensor contains only one component, and the coarse-graining procedure is exact in this case. Although the data aligns very well with the expected values even in the weak coupling regime for the average plaquette, this alignment is entirely coincidental. As will be shown in Section \ref{subsection:polyakov}, including only the trivial representation cannot correctly represent the physics of more complex observables.

The observed fluctuations for SU(2) at weak couplings can be attributed to the limitations of the character expansion. Away from the strong coupling regime, the character expansion becomes less accurate, particularly beyond the region where the strong coupling expansion diverges from the correct values. For SU(3), this divergence occurs beyond $\beta = 4$, which explains why similar fluctuations are not observed in figure \ref{fig:wilson} (right).

\subsection{Polyakov loop}
\label{subsection:polyakov}

\begin{figure}
    \centering
    \includegraphics[scale=0.8]{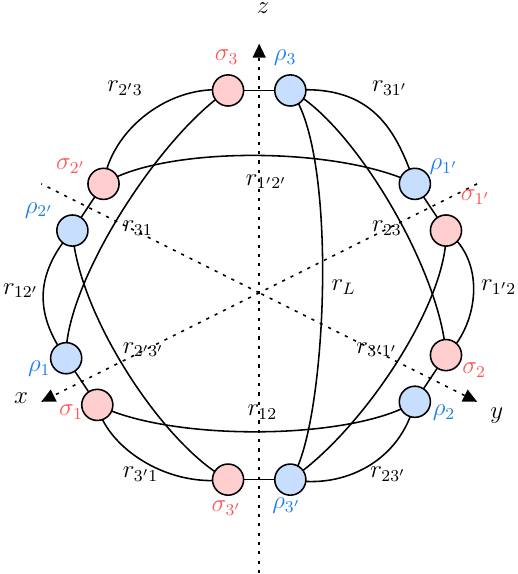}
    \caption{The armillary sphere corresponding to the extended action \eqref{eq:extended_action}. The additional index $r_L$ comes from the temporal link variables associated with the Polyakov loop.}
    \label{fig:armillary_polyakov}
\end{figure}

Apart from the fundamental Wilson loop; i.e., the plaquette \eqref{eq:plaquette}, we also consider the Polyakov loop operator
\begin{equation}
    L(x)= U_\tau(x)U_\tau(x+\hat\tau)\cdots U_\tau(x+(N_\tau-1)\hat\tau)
\end{equation}
which serves as the order parameter for the deconfinement transition. Here, $x$ is the lattice position on the temporal cross-section $\Lambda_2\subset\Lambda_3$ and $N_\tau$ is the size of the lattice in the imaginary time direction. To compute the expectation value of this observable, we introduce the Polyakov term to the action
\begin{align}
    S&=S_P+S_L;\label{eq:extended_action}\\
    S_L&=\frac{\kappa}{N}\sum_{x\in\Lambda_2}\Re\tr (\mathbb{1}-L(x)).\label{eq:polyakov_term}
\end{align}
The character expansion of the Polyakov term is
\begin{equation}
    e^{\frac{\kappa}{N}\Re\tr(L-\mathbb{1})}=\sum_rf_r(\kappa)\tr_rL.
\end{equation}
The expansion coefficient $f_r(\kappa)$ is given in \eqref{eq:expansion_coefficient}. We again separate the link variables in $\tr_r L$ and integrate them with those from the plaquette action. This gives an armillary sphere with a slightly different structure, as shown in figure \ref{fig:armillary_polyakov}. In this diagram, the $\pm z$ directions involve a 2-legged and a 3-legged\footnote{The 3-legged CG coefficient is a generalized Clebsch-Gordan coefficient obtained from integrating a product of 3 link variables. See appendix \ref{sec:gCGs} for its definition.} CG coefficient while the other directions only involve 2 2-legged coefficients. The corresponding 5-representation indices are
\begin{align}
    \tilde\alpha_z &= (r_{23},r_{31'},r_L;r_{2'3},r_{31};\rho_3,\sigma_3),\\
    \tilde\alpha_{z'} &= (r_{23'},r_{3'1'},r_L;r_{2'3'},r_{3'1};\rho_{3'},\sigma_{3'}).
\end{align}
The lists of all 5-representation indices for both SU(2) and SU(3) are given in appendix \ref{sec:mri}.

In addition to the expansion coefficients from the plaquette action, we also need to multiply another factor from the Polyakov action to the site tensor
\begin{equation}
    T_{\alpha_x\alpha_{x'}\alpha_y\alpha_{y'}\tilde\alpha_z\tilde\alpha_{z'}}=A_{\alpha_x\alpha_{x'}\alpha_y\alpha_{y'}\tilde\alpha_z\tilde\alpha_{z'}}
    \times\frac{(f_{r_{12}}(\beta)f_{r_{1'2}}(\beta)\cdots f_{r_{3'1'}}(\beta))^{1/4}}{\sqrt{
    d_{\rho_{1}}d_{\rho_{2}}d_{\rho_{3}}
    d_{\rho_{1'}}d_{\rho_{2'}}d_{\rho_{3'}}
    }}\times f_{r_L}^{1/N_\tau}(\kappa).
\end{equation}

In this work, we compute the Polyakov loop expectation value and its susceptibility via the first and second derivatives of $\log Z$ with respect to $\kappa$:
\begin{align}
    \bar L&=\frac{1}{N}\langle \Re\tr L\rangle = 1+\frac{N_\tau}{V}\frac{\partial}{\partial\kappa}\log Z(\beta,\kappa),
    \label{eq:polyakov_loop}\\
    \chi&=\frac{V}{N^2N_\tau}\left(\langle(\Re\tr L)^2\rangle-\langle\Re\tr L\rangle^2\right)
    =\frac{N_\tau}{V}\frac{\partial^2}{\partial\kappa^2}\log Z(\beta,\kappa);\label{eq:polyakov_susceptibility}
\end{align}
with
\begin{align}
    \frac{\partial}{\partial\kappa}\log Z(\beta,\kappa)
    &\approx \left.\frac{1}{2\epsilon}(\log Z(\beta,\kappa+\epsilon)-\log Z(\beta,\kappa-\epsilon))\right|_{\epsilon=0.01},\label{eq:first_diff}\\
    \frac{\partial^2}{\partial\kappa^2}\log Z(\beta,\kappa)
    &\approx \left.\frac{1}{\epsilon^2}(\log Z(\beta,\kappa+\epsilon)+\log Z(\beta,\kappa-\epsilon)-2\log Z(\beta,\kappa))\right|_{\epsilon=0.01}.\label{eq:second_diff}
\end{align}
If more accuracy is required, the numerical derivatives could be replaced by methods based on impurity tensors (See Ref.~\cite{Akiyama:2024qgv,Morita:2018tpw}, for example). However, such techniques are not needed in our parameter regions.

To induce the spontaneous symmetry breaking of the $\mathbb{Z}_N$ center symmetry, we compute $\bar L$ and $\chi$ at $\kappa=0.01$ and $V=1024^2\times 1$ in every calculation below.
We have confirmed that $\kappa=0.01$ is sufficiently small to be consistent with the $\kappa=0$ in the confined phase but is not too small to observe the finite-volume effect in the deconfined phase.
For $\text{SU}(2)$ gauge theory we perform the computation with only two terms in the Polyakov loop character expansion; $r_L\in\{\mathbf{1},\mathbf{2}\}$, while for $\text{SU}(3)$, we consider three terms; $r_L\in\{\mathbf{1},\mathbf{3},\bar{\mathbf{3}}\}$, since the contribution to the partition function from $\mathbf{3}$ and $\bar{\mathbf{3}}$ are equal.

\begin{figure}
    \centering
    \includegraphics[scale=0.65]{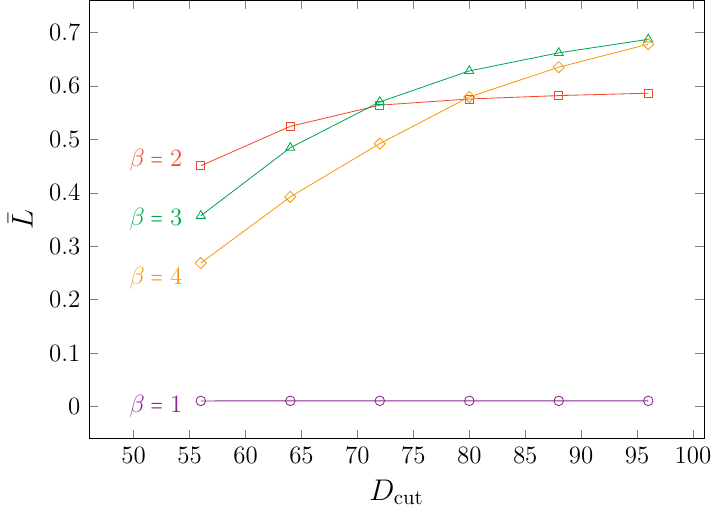}
    \includegraphics[scale=0.65]{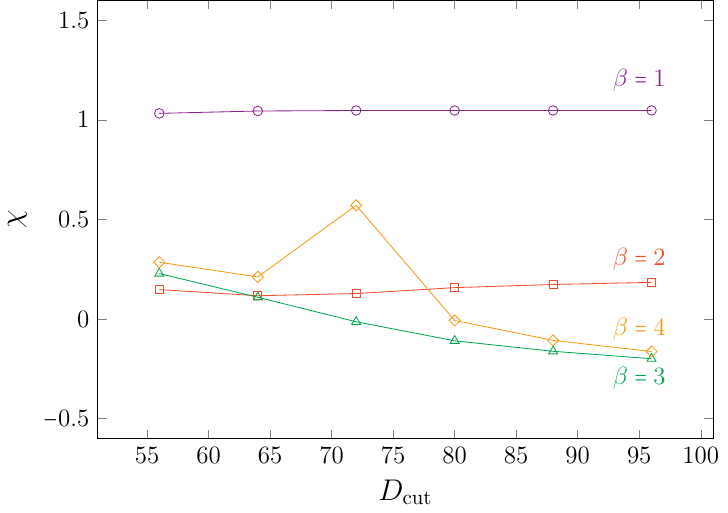}
    \caption{The $D_\text{cut}$ dependence of the Polyakov loop (left) and the susceptibility (right) of the $\text{SU}(2)$ gauge theory up to $D_\text{cut}=96$.}
    \label{fig:polyakob_dcut}
\end{figure}

\begin{figure}
    \centering
    \includegraphics[scale=0.65]{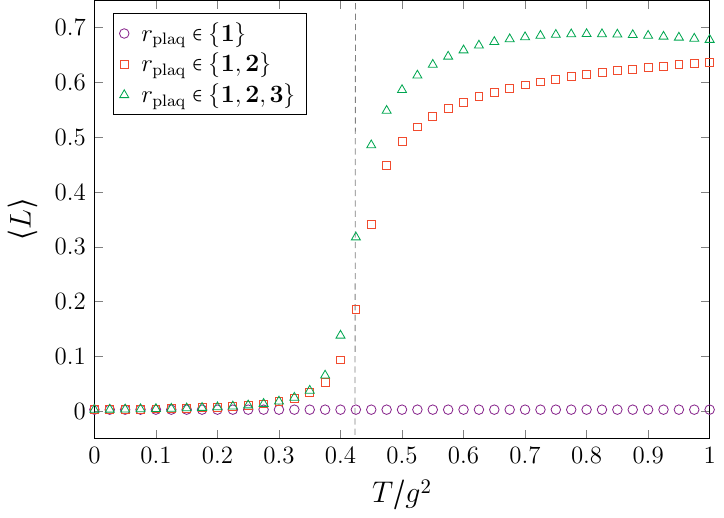}
    \includegraphics[scale=0.65]{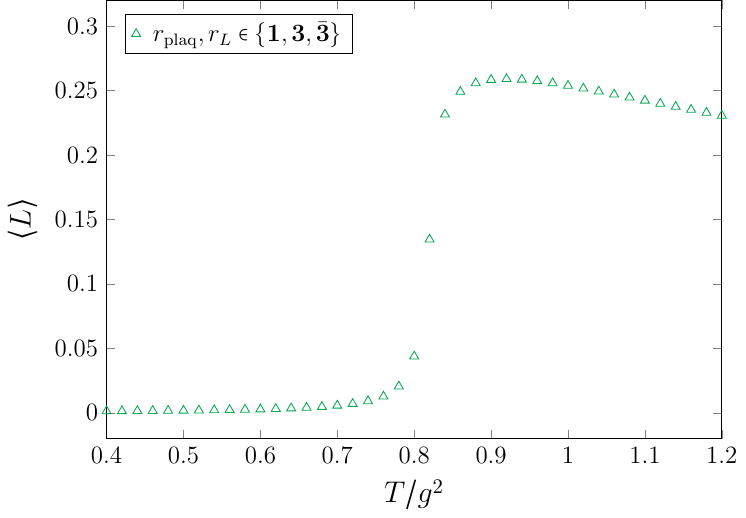}
    \caption{Polyakov loop \eqref{eq:polyakov_loop} as a function of temperature $T/g^2\equiv \beta/2NN_\tau$ with $N_\tau=1$ for $\text{SU}(2)$ (left) and $\text{SU}(3)$ (right) gauge theories. The vertical line shows the naive estimate of the critical temperature for $\text{SU}(2)$ from Ref. \cite{Teper:1993gp}.}
    \label{fig:polyakov_loop_su2}
\end{figure}

\begin{figure}
    \centering
    \includegraphics[scale=0.65]{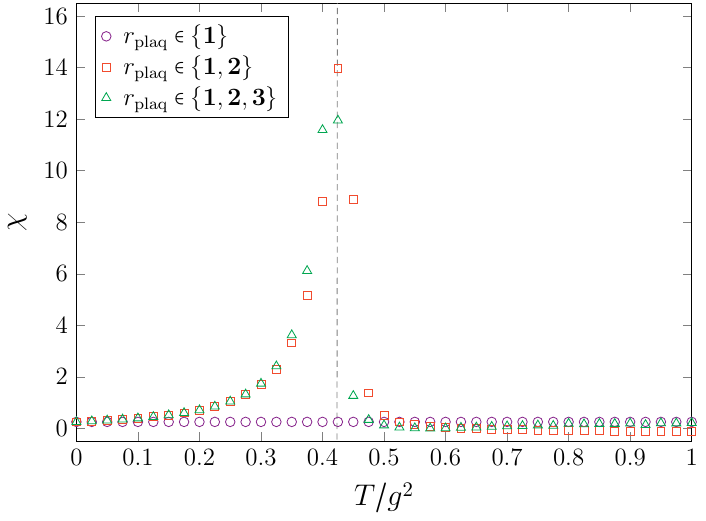}
    \includegraphics[scale=0.65]{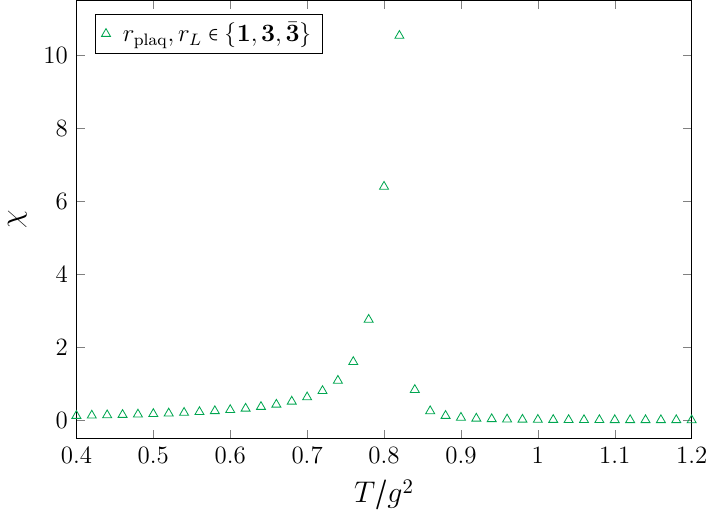}
    \caption{Polyakov susceptibility \eqref{eq:polyakov_susceptibility} as a function of temperature $T/g^2\equiv \beta/2NN_\tau$ with $N_\tau=1$ for $\text{SU}(2)$ (left) and $\text{SU}(3)$ (right) gauge theories. The vertical line shows the naive estimate of the critical temperature for $\text{SU}(2)$.}
    \label{fig:polyakov_su2}
\end{figure}

To observe the deconfinement transition, we perform the calculation at various temperatures $T/g^2\equiv \beta/2NN_\tau$. In order to access sufficiently high-temperature regions, we consider $N_\tau=1$. We therefore do not need to perform a coarse-graining procedure in the temporal direction. Coarse-graining in the spatial plane is done using the Levin-Nave TRG algorithm \cite{Levin:2006jai}. Figure \ref{fig:polyakob_dcut} shows the $D_\text{cut}$ dependence of the Polyakov loop and its susceptibility ($N=2$) with $\beta=1,2,3,4$.
We find that the convergence becomes significantly slower as we increase $\beta$ beyond $\beta=2$.
This signifies the transition from the confined phase to the deconfined phase as we will see next. In the following computations, we use $D_\text{cut}=96$ for both the compression and the coarse-graining procedure. The result is shown in figure \ref{fig:polyakov_loop_su2} and \ref{fig:polyakov_su2}.

For $\text{SU}(2)$, the Polyakov loop susceptibility shows weak dependence on the number of Wilson character expansion terms \eqref{eq:character_expansion} in the confined phase $T/g^2\lesssim0.4$. However, the truncation effect is more prominent as we approach the deconfined phase. In the deconfined region, more expansion terms and larger bond dimensions are needed to obtain more precise values.
Despite that, our technique can be used to qualitatively identify the deconfinement temperature, which is close to the naive estimate\footnote{Obtained by extrapolating the critical coupling $\beta_c$ to $N_\tau=1$ with the constraint $\beta_c=0$ at $N_\tau=0$.} of $T_c/g^2=\beta_c/4=0.4241$ from Ref. \cite{Teper:1993gp}. The same computation is also done for $\text{SU}(3)$, which shows the deconfinement transition at around $T_c/g^2=\beta_c/6\approx0.81$. Similar to SU(2), a larger number of expansion terms and bond dimensions are required to investigate the deconfined phase more quantitatively.

\section{Summary and discussions}
\label{section:summary}
In this paper, we perform the tensor renormalization group computation of the three-dimensional pure $\text{SU}(2)$ and $\text{SU}(3)$ gauge theories using the reduced tensor network formulation proposed in Ref.\cite{Yosprakob:2023jgl}. The formulation uses the fact that the group integral of the link variables can be done analytically and gives a tensor network where matrix indices can be traced out completely. These matrix indices, associated with internal symmetry, typically result in a large degeneracy of the singular value spectrum in the TRG computation. To demonstrate the usefulness of this technique, we compute the average plaquette at zero temperature. The result agrees very well with the strong-coupling expansions for both $\text{SU}(2)$ and $\text{SU}(3)$. We also reproduce the deconfinement transition and identify the transition temperature, which is in agreement with the Monte Carlo result for $\text{SU}(2)$.

We observe that the average plaquette is less sensitive to bond dimensions and expansion terms than the Polyakov loop. In the deconfined region, a large number of expansion terms and bond dimensions are required to obtain a precise value of Polyakov loop-related quantities. Such a strong dependence on the bond dimension is a common feature of topological observables. Another possibility is that, at high temperatures, the theory with $N_\tau=1$ may exhibit similar behavior as the quasi-ordered phase of the Berezinskii–Kosterlitz–Thouless (BKT) transition for a planar lattice system. In this phase, the correlation function decays like a power law and thus requires large bond dimensions in the tensor network. It would be fruitful to investigate this behavior in four dimensions, where the BKT transition disappears.

It should be noted that, to properly study the deconfinement transition beyond identifying the critical temperature, the infinite-volume limit must be taken first, followed by the $\kappa \rightarrow 0$ limit, in this specific order. Thus, both $\kappa$ and $\epsilon$ in \eqref{eq:first_diff} and \eqref{eq:second_diff} must be fine-tuned as we consider larger volumes. We plan to address this in future work.

For future directions, 
this technique could be combined with the multi-layer construction \cite{Yosprakob:2023tyr, Akiyama:2023lvr} to simulate a QCD-like theory. In four dimensions, one could also construct an armillary sphere associated with a gauge theory with a $\theta$ term, which suffers from a severe sign problem, especially at $\theta=\pi$. In this direction, one can study the relation between different $\theta$-vacuum topological sectors and the representations of the gauge group. Such a relationship was found to exist in the two-dimensional $\text{U}(N)$ theory \cite{Hirasawa:2021qvh}. By clarifying such a relation in four dimensions, we could deepen the understanding of the $\theta$ term.

\section*{Acknowledgments}
We thank Shinichiro Akiyama, Akira Matsumoto, and Jun Nishimura for their valuable discussions.
This work is supported by a Grant-in-Aid for Transformative Research Areas “The Natural Laws of Extreme Universe—A New Paradigm for Spacetime and Matter from Quantum Information” (KAKENHI Grant No. JP21H05191) from JSPS of Japan, and JST CREST Grant Number JPMJCR24I1.

\appendix

\section{Generalized Clebsch-Gordan coefficients}
\label{sec:gCGs}
In this section, we give a brief review of the generalized Clebsch-Gordan coefficients that arise from integrating the link variables. Here, we only consider the 3-legged CG coefficients. More general cases are discussed in Ref.~\cite{Yosprakob:2023jgl}.

We consider a decomposition of the product of 3 group elements into irreducible components
\begin{equation}
    U^{r_1}_{i_1i'_1}U^{r_2}_{i_2i'_2}U^{r_3}_{i_3i'_3}
    =\sum_{\rho\in D_{r_1\otimes r_2\otimes r_3}}\sum_{\hat\imath,\hat\imath'} C^{\rho\hat\imath}_{r_1i_1;r_2i_2;r_3i_3}C^{\rho\hat\imath'}_{r_1i'_1;r_2i'_2;r_3i'_3}U^\rho_{\hat\imath\hat\imath'}
    \label{eq:3leggedCGdecomposition}
\end{equation}
where $D_{r_1\otimes r_2\otimes r_3}$ denotes a set of irreducible components of the product $r_1\otimes r_2\otimes r_3$. The 3-legged CG coefficients $C^{\rho\hat\imath}_{r_1i_1;r_2i_2;r_3i_3}$ can be considered as an orthogonal transformation matrix that relates the uncoupled basis $\{r_1,r_2,r_3\}$ to the decomposed basis $\{\rho\in D_{r_1\otimes r_2\otimes r_3}\}$. It can be obtained by first decomposing the product of $r_1$ and $r_2$ (see \eqref{eq:cgdecomposition}) and then decomposing the product of its result and $r_3$:

\begin{align}
    \left(U^{r_1}_{i_1i'_1}\right.&\left.U^{r_2}_{i_2i'_2}\right)U^{r_3}_{i_3i'_3}
    =\left(
    \sum_{\sigma\in D_{r_1\otimes r_2}}
    \sum_{\tilde\imath,\tilde\imath'}
    C^{\sigma\tilde\imath}_{r_1i_1;r_2i_2}
    C^{\sigma\tilde\imath'}_{r_1i'_1;r_2i'_2}
    U^{\sigma}_{\tilde\imath\tilde\imath'}
    \right)U^{r_3}_{i_3i'_3}\\
    &=
    \sum_{\sigma\in D_{r_1\otimes r_2}}
    \sum_{\tilde\imath,\tilde\imath'}
    C^{\sigma\tilde\imath}_{r_1i_1;r_2i_2}
    C^{\sigma\tilde\imath'}_{r_1i'_1;r_2i'_2}
    \left(
    U^{\sigma}_{\tilde\imath\tilde\imath'}U^{r_3}_{i_3i'_3}
    \right)\\
    &=
    \sum_{\sigma\in D_{r_1\otimes r_2}}
    \sum_{\tilde\imath,\tilde\imath'}
    C^{\sigma\tilde\imath}_{r_1i_1;r_2i_2}
    C^{\sigma\tilde\imath'}_{r_1i'_1;r_2i'_2}
    \sum_{\rho\in D_{\sigma\otimes r_3}}
    \sum_{\hat\imath,\hat\imath'}
    C^{\rho\hat\imath}_{\sigma\tilde\imath;r_3i_3}
    C^{\rho\hat\imath'}_{\sigma\tilde\imath';r_3i'_3}U^\rho_{\hat\imath\hat\imath'}\\
    &=
    \sum_{\sigma\in D_{r_1\otimes r_2}}
    \sum_{\rho\in D_{\sigma\otimes r_3}}
    \sum_{\hat\imath,\hat\imath'}
    \left(
    \sum_{\tilde\imath}
    C^{\sigma\tilde\imath}_{r_1i_1;r_2i_2}
    C^{\rho\hat\imath}_{\sigma\tilde\imath;r_3i_3}\right)
    \left(
    \sum_{\tilde\imath'}
    C^{\sigma\tilde\imath'}_{r_1i'_1;r_2i'_2}
    C^{\rho\hat\imath'}_{\sigma\tilde\imath';r_3i'_3}\right)U^\rho_{\hat\imath\hat\imath'}.
\end{align}

Defining
\begin{equation}
    \sum_{\rho\in D_{r_1\otimes r_2\otimes r_3}}\equiv\sum_{\sigma\in D_{r_1\otimes r_2}}
    \sum_{\rho\in D_{\sigma\otimes r_3}},
\end{equation}
we can identify the 3-legged CG coefficient in \eqref{eq:3leggedCGdecomposition} as
\begin{equation}
    C^{\rho\hat\imath}_{r_1i_1;r_2i_2;r_3i_3}\equiv \sum_{\tilde\imath}C^{\sigma\tilde\imath}_{r_1i_1;r_2i_2}C^{\rho\hat\imath}_{\sigma\tilde\imath;r_3i_3}.
    \label{eq:3leggedCG}
\end{equation}

\section{Multi-representation indices}
\label{sec:mri}

In this section, we list the multi-representation indices that serve as the tensor legs of the armillary spheres in different setups.
\subsection{4-representation indices}
Consider a decomposition of the tensor product $r_1\otimes r_2$ into irreducible components. Let $\rho$ be one of such components. Similarly, let $\sigma$ be an irreducible component of the tensor product $s_1\otimes s_2$. If $\rho\simeq\sigma$, we denote a 4-representation index for this case as
\begin{equation}
    (r_1, r_2; s_1, s_2; \rho,\sigma)
    \label{eq:su2_pure}
\end{equation}
Table \ref{tab:su2_pure} shows the 4-representation indices for SU(2) where $r_1, r_2, s_1, s_2$ are $\mathbf{1}$, $\mathbf{2}$, or $\mathbf{3}$. Table \ref{tab:su3_pure} shows the 4-representation indices for SU(3) where $r_1, r_2, s_1, s_2$ are $\mathbf{1}$, $\mathbf{3}$, or $\bar{\mathbf{3}}$. The representation $\rho$ and $\sigma$ are not restricted.
\begin{table}
    \centering
    {\footnotesize
    \begin{tabular}{|c|c|}
    \hline
        index & symbol\\\hline
        1 & $(\textbf{1}, \textbf{1}; \textbf{1}, \textbf{1}; \textbf{1}, \textbf{1})$ \\
        2 & $(\textbf{1}, \textbf{1}; \textbf{2}, \textbf{2}; \textbf{1}, \textbf{1})$ \\
        3 & $(\textbf{1}, \textbf{1}; \textbf{3}, \textbf{3}; \textbf{1}, \textbf{1})$ \\
        4 & $(\textbf{1}, \textbf{2}; \textbf{1}, \textbf{2}; \textbf{2}, \textbf{2})$ \\
        5 & $(\textbf{1}, \textbf{2}; \textbf{2}, \textbf{1}; \textbf{2}, \textbf{2})$ \\
        6 & $(\textbf{1}, \textbf{2}; \textbf{2}, \textbf{3}; \textbf{2}, \textbf{2})$ \\
        7 & $(\textbf{1}, \textbf{2}; \textbf{3}, \textbf{2}; \textbf{2}, \textbf{2})$ \\
        8 & $(\textbf{1}, \textbf{3}; \textbf{1}, \textbf{3}; \textbf{3}, \textbf{3})$ \\
        9 & $(\textbf{1}, \textbf{3}; \textbf{2}, \textbf{2}; \textbf{3}, \textbf{3})$ \\
        10 & $(\textbf{1}, \textbf{3}; \textbf{3}, \textbf{1}; \textbf{3}, \textbf{3})$ \\
        11 & $(\textbf{1}, \textbf{3}; \textbf{3}, \textbf{3}; \textbf{3}, \textbf{3})$ \\
        12 & $(\textbf{2}, \textbf{1}; \textbf{1}, \textbf{2}; \textbf{2}, \textbf{2})$ \\
        \hline
    \end{tabular}
    \begin{tabular}{|c|c|}
    \hline
         & symbol\\\hline
        13 & $(\textbf{2}, \textbf{1}; \textbf{2}, \textbf{1}; \textbf{2}, \textbf{2})$ \\
        14 & $(\textbf{2}, \textbf{1}; \textbf{2}, \textbf{3}; \textbf{2}, \textbf{2})$ \\
        15 & $(\textbf{2}, \textbf{1}; \textbf{3}, \textbf{2}; \textbf{2}, \textbf{2})$ \\
        16 & $(\textbf{2}, \textbf{2}; \textbf{1}, \textbf{1}; \textbf{1}, \textbf{1})$ \\
        17 & $(\textbf{2}, \textbf{2}; \textbf{1}, \textbf{3}; \textbf{3}, \textbf{3})$ \\
        18 & $(\textbf{2}, \textbf{2}; \textbf{2}, \textbf{2}; \textbf{1}, \textbf{1})$ \\
        19 & $(\textbf{2}, \textbf{2}; \textbf{2}, \textbf{2}; \textbf{3}, \textbf{3})$ \\
        20 & $(\textbf{2}, \textbf{2}; \textbf{3}, \textbf{1}; \textbf{3}, \textbf{3})$ \\
        21 & $(\textbf{2}, \textbf{2}; \textbf{3}, \textbf{3}; \textbf{1}, \textbf{1})$ \\
        22 & $(\textbf{2}, \textbf{2}; \textbf{3}, \textbf{3}; \textbf{3}, \textbf{3})$ \\
        23 & $(\textbf{2}, \textbf{3}; \textbf{1}, \textbf{2}; \textbf{2}, \textbf{2})$ \\
        24 & $(\textbf{2}, \textbf{3}; \textbf{2}, \textbf{1}; \textbf{2}, \textbf{2})$ \\
        \hline
    \end{tabular}
    \begin{tabular}{|c|c|}
    \hline
         & symbol\\\hline
        25 & $(\textbf{2}, \textbf{3}; \textbf{2}, \textbf{3}; \textbf{2}, \textbf{2})$ \\
        26 & $(\textbf{2}, \textbf{3}; \textbf{2}, \textbf{3}; \textbf{4}, \textbf{4})$ \\
        27 & $(\textbf{2}, \textbf{3}; \textbf{3}, \textbf{2}; \textbf{2}, \textbf{2})$ \\
        28 & $(\textbf{2}, \textbf{3}; \textbf{3}, \textbf{2}; \textbf{4}, \textbf{4})$ \\
        29 & $(\textbf{3}, \textbf{1}; \textbf{1}, \textbf{3}; \textbf{3}, \textbf{3})$ \\
        30 & $(\textbf{3}, \textbf{1}; \textbf{2}, \textbf{2}; \textbf{3}, \textbf{3})$ \\
        31 & $(\textbf{3}, \textbf{1}; \textbf{3}, \textbf{1}; \textbf{3}, \textbf{3})$ \\
        32 & $(\textbf{3}, \textbf{1}; \textbf{3}, \textbf{3}; \textbf{3}, \textbf{3})$ \\
        33 & $(\textbf{3}, \textbf{2}; \textbf{1}, \textbf{2}; \textbf{2}, \textbf{2})$ \\
        34 & $(\textbf{3}, \textbf{2}; \textbf{2}, \textbf{1}; \textbf{2}, \textbf{2})$ \\
        35 & $(\textbf{3}, \textbf{2}; \textbf{2}, \textbf{3}; \textbf{2}, \textbf{2})$ \\
        36 & $(\textbf{3}, \textbf{2}; \textbf{2}, \textbf{3}; \textbf{4}, \textbf{4})$ \\
        \hline
    \end{tabular}
    \begin{tabular}{|c|c|}
    \hline
         & symbol\\\hline
        37 & $(\textbf{3}, \textbf{2}; \textbf{3}, \textbf{2}; \textbf{2}, \textbf{2})$ \\
        38 & $(\textbf{3}, \textbf{2}; \textbf{3}, \textbf{2}; \textbf{4}, \textbf{4})$ \\
        39 & $(\textbf{3}, \textbf{3}; \textbf{1}, \textbf{1}; \textbf{1}, \textbf{1})$ \\
        40 & $(\textbf{3}, \textbf{3}; \textbf{1}, \textbf{3}; \textbf{3}, \textbf{3})$ \\
        41 & $(\textbf{3}, \textbf{3}; \textbf{2}, \textbf{2}; \textbf{1}, \textbf{1})$ \\
        42 & $(\textbf{3}, \textbf{3}; \textbf{2}, \textbf{2}; \textbf{3}, \textbf{3})$ \\
        43 & $(\textbf{3}, \textbf{3}; \textbf{3}, \textbf{1}; \textbf{3}, \textbf{3})$ \\
        44 & $(\textbf{3}, \textbf{3}; \textbf{3}, \textbf{3}; \textbf{1}, \textbf{1})$ \\
        45 & $(\textbf{3}, \textbf{3}; \textbf{3}, \textbf{3}; \textbf{3}, \textbf{3})$ \\
        46 & $(\textbf{3}, \textbf{3}; \textbf{3}, \textbf{3}; \textbf{5}, \textbf{5})$ \\
        &\\
        &\\
        \hline
    \end{tabular}
    }
    \caption{4-representation indices for SU(2).}
    \label{tab:su2_pure}
\end{table}
\begin{table}
    \centering
    {\footnotesize
    \begin{tabular}{|c|c|}
    \hline
        index & symbol\\\hline
        1 & $(\textbf{1}, \textbf{1}; \textbf{1}, \textbf{1}; \textbf{1}, \textbf{1})$ \\
        2 & $(\textbf{1}, \textbf{1}; \textbf{3}, \bar{\textbf{3}}; \textbf{1}, \textbf{1})$ \\
        3 & $(\textbf{1}, \textbf{1}; \bar{\textbf{3}}, \textbf{3}; \textbf{1}, \textbf{1})$ \\
        4 & $(\textbf{1}, \textbf{3}; \textbf{1}, \textbf{3}; \textbf{3}, \textbf{3})$ \\
        5 & $(\textbf{1}, \textbf{3}; \textbf{3}, \textbf{1}; \textbf{3}, \textbf{3})$ \\
        6 & $(\textbf{1}, \textbf{3}; \bar{\textbf{3}}, \bar{\textbf{3}}; \textbf{3}, \textbf{3})$ \\
        7 & $(\textbf{1}, \bar{\textbf{3}}; \textbf{1}, \bar{\textbf{3}}; \bar{\textbf{3}}, \bar{\textbf{3}})$ \\
        8 & $(\textbf{1}, \bar{\textbf{3}}; \textbf{3}, \textbf{3}; \bar{\textbf{3}}, \bar{\textbf{3}})$ \\
        9 & $(\textbf{1}, \bar{\textbf{3}}; \bar{\textbf{3}}, \textbf{1}; \bar{\textbf{3}}, \bar{\textbf{3}})$ \\
        10 & $(\textbf{3}, \textbf{1}; \textbf{1}, \textbf{3}; \textbf{3}, \textbf{3})$ \\
        11 & $(\textbf{3}, \textbf{1}; \textbf{3}, \textbf{1}; \textbf{3}, \textbf{3})$ \\
        \hline
    \end{tabular}
    \begin{tabular}{|c|c|}
    \hline
         & symbol\\\hline
        12 & $(\textbf{3}, \textbf{1}; \bar{\textbf{3}}, \bar{\textbf{3}}; \textbf{3}, \textbf{3})$ \\
        13 & $(\textbf{3}, \textbf{3}; \textbf{1}, \bar{\textbf{3}}; \bar{\textbf{3}}, \bar{\textbf{3}})$ \\
        14 & $(\textbf{3}, \textbf{3}; \textbf{3}, \textbf{3}; \bar{\textbf{3}}, \bar{\textbf{3}})$ \\
        15 & $(\textbf{3}, \textbf{3}; \textbf{3}, \textbf{3}; \textbf{6}, \textbf{6})$ \\
        16 & $(\textbf{3}, \textbf{3}; \bar{\textbf{3}}, \textbf{1}; \bar{\textbf{3}}, \bar{\textbf{3}})$ \\
        17 & $(\textbf{3}, \bar{\textbf{3}}; \textbf{1}, \textbf{1}; \textbf{1}, \textbf{1})$ \\
        18 & $(\textbf{3}, \bar{\textbf{3}}; \textbf{3}, \bar{\textbf{3}}; \textbf{1}, \textbf{1})$ \\
        19 & $(\textbf{3}, \bar{\textbf{3}}; \textbf{3}, \bar{\textbf{3}}; \textbf{8}, \textbf{8})$ \\
        20 & $(\textbf{3}, \bar{\textbf{3}}; \bar{\textbf{3}}, \textbf{3}; \textbf{1}, \textbf{1})$ \\
        21 & $(\textbf{3}, \bar{\textbf{3}}; \bar{\textbf{3}}, \textbf{3}; \textbf{8}, \textbf{8})$ \\
        22 & $(\bar{\textbf{3}}, \textbf{1}; \textbf{1}, \bar{\textbf{3}}; \bar{\textbf{3}}, \bar{\textbf{3}})$ \\
        \hline
    \end{tabular}
    \begin{tabular}{|c|c|}
    \hline
         & symbol\\\hline
        23 & $(\bar{\textbf{3}}, \textbf{1}; \textbf{3}, \textbf{3}; \bar{\textbf{3}}, \bar{\textbf{3}})$ \\
        24 & $(\bar{\textbf{3}}, \textbf{1}; \bar{\textbf{3}}, \textbf{1}; \bar{\textbf{3}}, \bar{\textbf{3}})$ \\
        25 & $(\bar{\textbf{3}}, \textbf{3}; \textbf{1}, \textbf{1}; \textbf{1}, \textbf{1})$ \\
        26 & $(\bar{\textbf{3}}, \textbf{3}; \textbf{3}, \bar{\textbf{3}}; \textbf{1}, \textbf{1})$ \\
        27 & $(\bar{\textbf{3}}, \textbf{3}; \textbf{3}, \bar{\textbf{3}}; \textbf{8}, \textbf{8})$ \\
        28 & $(\bar{\textbf{3}}, \textbf{3}; \bar{\textbf{3}}, \textbf{3}; \textbf{1}, \textbf{1})$ \\
        29 & $(\bar{\textbf{3}}, \textbf{3}; \bar{\textbf{3}}, \textbf{3}; \textbf{8}, \textbf{8})$ \\
        30 & $(\bar{\textbf{3}}, \bar{\textbf{3}}; \textbf{1}, \textbf{3}; \textbf{3}, \textbf{3})$ \\
        31 & $(\bar{\textbf{3}}, \bar{\textbf{3}}; \textbf{3}, \textbf{1}; \textbf{3}, \textbf{3})$ \\
        32 & $(\bar{\textbf{3}}, \bar{\textbf{3}}; \bar{\textbf{3}}, \bar{\textbf{3}}; \textbf{3}, \textbf{3})$ \\
        33 & $(\bar{\textbf{3}}, \bar{\textbf{3}}; \bar{\textbf{3}}, \bar{\textbf{3}}; \bar{\textbf{6}}, \bar{\textbf{6}})$ \\
        \hline
    \end{tabular}
    }
    \caption{4-representation indices for SU(3).}
    \label{tab:su3_pure}
\end{table}

\subsection{5-representation indices}
Similar to 4-representation indices, 5-representation indices are associated with matching the decompositions of $r_1\otimes r_2\otimes r_3$ and $s_1\otimes s_2$. Namely, let $\rho$ be an irreducible component of $r_1\otimes r_2\otimes r_3$ and $\sigma$ be an irreducible component of $s_1\otimes s_2$. If $\rho\simeq \sigma$, we denote the associated 5-representation index as
\begin{equation}
    (r_1, r_2, r_3; s_1, s_2; \rho,\sigma)
    \label{eq:su2_poly}
\end{equation}

Table \ref{tab:su2_poly} shows the 5-representation indices for SU(2) where $r_1, r_2, r_3, s_1, s_2$ are $\mathbf{1}$, $\mathbf{2}$, or $\mathbf{3}$. Table \ref{tab:su3_poly} shows the 5-representation indices for SU(3) where $r_1, r_2, r_3, s_1, s_2$ are $\mathbf{1}$, $\mathbf{3}$, or $\bar{\mathbf{3}}$. The representation $\rho$ and $\sigma$ are not restricted.

\begin{table}
    \centering
    {\scriptsize
    \begin{tabular}{|c|c|}
    \hline
        index & symbol\\\hline
        1 & $(\textbf{1}, \textbf{1}, \textbf{1}; \textbf{1}, \textbf{1}; \textbf{1}, \textbf{1})$ \\
        2 & $(\textbf{1}, \textbf{1}, \textbf{1}; \textbf{2}, \textbf{2}; \textbf{1}, \textbf{1})$ \\
        3 & $(\textbf{1}, \textbf{1}, \textbf{1}; \textbf{3}, \textbf{3}; \textbf{1}, \textbf{1})$ \\
        4 & $(\textbf{1}, \textbf{1}, \textbf{2}; \textbf{1}, \textbf{2}; \textbf{2}, \textbf{2})$ \\
        5 & $(\textbf{1}, \textbf{1}, \textbf{2}; \textbf{2}, \textbf{1}; \textbf{2}, \textbf{2})$ \\
        6 & $(\textbf{1}, \textbf{1}, \textbf{2}; \textbf{2}, \textbf{3}; \textbf{2}, \textbf{2})$ \\
        7 & $(\textbf{1}, \textbf{1}, \textbf{2}; \textbf{3}, \textbf{2}; \textbf{2}, \textbf{2})$ \\
        8 & $(\textbf{1}, \textbf{2}, \textbf{1}; \textbf{1}, \textbf{2}; \textbf{2}, \textbf{2})$ \\
        9 & $(\textbf{1}, \textbf{2}, \textbf{1}; \textbf{2}, \textbf{1}; \textbf{2}, \textbf{2})$ \\
        10 & $(\textbf{1}, \textbf{2}, \textbf{1}; \textbf{2}, \textbf{3}; \textbf{2}, \textbf{2})$ \\
        11 & $(\textbf{1}, \textbf{2}, \textbf{1}; \textbf{3}, \textbf{2}; \textbf{2}, \textbf{2})$ \\
        12 & $(\textbf{1}, \textbf{2}, \textbf{2}; \textbf{1}, \textbf{1}; \textbf{1}, \textbf{1})$ \\
        13 & $(\textbf{1}, \textbf{2}, \textbf{2}; \textbf{1}, \textbf{3}; \textbf{3}, \textbf{3})$ \\
        14 & $(\textbf{1}, \textbf{2}, \textbf{2}; \textbf{2}, \textbf{2}; \textbf{1}, \textbf{1})$ \\
        15 & $(\textbf{1}, \textbf{2}, \textbf{2}; \textbf{2}, \textbf{2}; \textbf{3}, \textbf{3})$ \\
        16 & $(\textbf{1}, \textbf{2}, \textbf{2}; \textbf{3}, \textbf{1}; \textbf{3}, \textbf{3})$ \\
        17 & $(\textbf{1}, \textbf{2}, \textbf{2}; \textbf{3}, \textbf{3}; \textbf{1}, \textbf{1})$ \\
        18 & $(\textbf{1}, \textbf{2}, \textbf{2}; \textbf{3}, \textbf{3}; \textbf{3}, \textbf{3})$ \\
        19 & $(\textbf{1}, \textbf{3}, \textbf{1}; \textbf{1}, \textbf{3}; \textbf{3}, \textbf{3})$ \\
        20 & $(\textbf{1}, \textbf{3}, \textbf{1}; \textbf{2}, \textbf{2}; \textbf{3}, \textbf{3})$ \\
        21 & $(\textbf{1}, \textbf{3}, \textbf{1}; \textbf{3}, \textbf{1}; \textbf{3}, \textbf{3})$ \\
        22 & $(\textbf{1}, \textbf{3}, \textbf{1}; \textbf{3}, \textbf{3}; \textbf{3}, \textbf{3})$ \\
        23 & $(\textbf{1}, \textbf{3}, \textbf{2}; \textbf{1}, \textbf{2}; \textbf{2}, \textbf{2})$ \\
        24 & $(\textbf{1}, \textbf{3}, \textbf{2}; \textbf{2}, \textbf{1}; \textbf{2}, \textbf{2})$ \\
        25 & $(\textbf{1}, \textbf{3}, \textbf{2}; \textbf{2}, \textbf{3}; \textbf{2}, \textbf{2})$ \\
        26 & $(\textbf{1}, \textbf{3}, \textbf{2}; \textbf{2}, \textbf{3}; \textbf{4}, \textbf{4})$ \\
        27 & $(\textbf{1}, \textbf{3}, \textbf{2}; \textbf{3}, \textbf{2}; \textbf{2}, \textbf{2})$ \\
        28 & $(\textbf{1}, \textbf{3}, \textbf{2}; \textbf{3}, \textbf{2}; \textbf{4}, \textbf{4})$ \\
        29 & $(\textbf{2}, \textbf{1}, \textbf{1}; \textbf{1}, \textbf{2}; \textbf{2}, \textbf{2})$ \\
        30 & $(\textbf{2}, \textbf{1}, \textbf{1}; \textbf{2}, \textbf{1}; \textbf{2}, \textbf{2})$ \\
        31 & $(\textbf{2}, \textbf{1}, \textbf{1}; \textbf{2}, \textbf{3}; \textbf{2}, \textbf{2})$ \\
        \hline
    \end{tabular}
    \begin{tabular}{|c|c|}
    \hline
         & symbol\\\hline
        32 & $(\textbf{2}, \textbf{1}, \textbf{1}; \textbf{3}, \textbf{2}; \textbf{2}, \textbf{2})$ \\
        33 & $(\textbf{2}, \textbf{1}, \textbf{2}; \textbf{1}, \textbf{1}; \textbf{1}, \textbf{1})$ \\
        34 & $(\textbf{2}, \textbf{1}, \textbf{2}; \textbf{1}, \textbf{3}; \textbf{3}, \textbf{3})$ \\
        35 & $(\textbf{2}, \textbf{1}, \textbf{2}; \textbf{2}, \textbf{2}; \textbf{1}, \textbf{1})$ \\
        36 & $(\textbf{2}, \textbf{1}, \textbf{2}; \textbf{2}, \textbf{2}; \textbf{3}, \textbf{3})$ \\
        37 & $(\textbf{2}, \textbf{1}, \textbf{2}; \textbf{3}, \textbf{1}; \textbf{3}, \textbf{3})$ \\
        38 & $(\textbf{2}, \textbf{1}, \textbf{2}; \textbf{3}, \textbf{3}; \textbf{1}, \textbf{1})$ \\
        39 & $(\textbf{2}, \textbf{1}, \textbf{2}; \textbf{3}, \textbf{3}; \textbf{3}, \textbf{3})$ \\
        40 & $(\textbf{2}, \textbf{2}, \textbf{1}; \textbf{1}, \textbf{1}; \textbf{1}, \textbf{1})$ \\
        41 & $(\textbf{2}, \textbf{2}, \textbf{1}; \textbf{1}, \textbf{3}; \textbf{3}, \textbf{3})$ \\
        42 & $(\textbf{2}, \textbf{2}, \textbf{1}; \textbf{2}, \textbf{2}; \textbf{1}, \textbf{1})$ \\
        43 & $(\textbf{2}, \textbf{2}, \textbf{1}; \textbf{2}, \textbf{2}; \textbf{3}, \textbf{3})$ \\
        44 & $(\textbf{2}, \textbf{2}, \textbf{1}; \textbf{3}, \textbf{1}; \textbf{3}, \textbf{3})$ \\
        45 & $(\textbf{2}, \textbf{2}, \textbf{1}; \textbf{3}, \textbf{3}; \textbf{1}, \textbf{1})$ \\
        46 & $(\textbf{2}, \textbf{2}, \textbf{1}; \textbf{3}, \textbf{3}; \textbf{3}, \textbf{3})$ \\
        47 & $(\textbf{2}, \textbf{2}, \textbf{2}; \textbf{1}, \textbf{2}; \textbf{2}_1, \textbf{2})$ \\
        48 & $(\textbf{2}, \textbf{2}, \textbf{2}; \textbf{1}, \textbf{2}; \textbf{2}_2, \textbf{2})$ \\
        49 & $(\textbf{2}, \textbf{2}, \textbf{2}; \textbf{2}, \textbf{1}; \textbf{2}_1, \textbf{2})$ \\
        50 & $(\textbf{2}, \textbf{2}, \textbf{2}; \textbf{2}, \textbf{1}; \textbf{2}_2, \textbf{2})$ \\
        51 & $(\textbf{2}, \textbf{2}, \textbf{2}; \textbf{2}, \textbf{3}; \textbf{2}_1, \textbf{2})$ \\
        52 & $(\textbf{2}, \textbf{2}, \textbf{2}; \textbf{2}, \textbf{3}; \textbf{2}_2, \textbf{2})$ \\
        53 & $(\textbf{2}, \textbf{2}, \textbf{2}; \textbf{2}, \textbf{3}; \textbf{4}, \textbf{4})$ \\
        54 & $(\textbf{2}, \textbf{2}, \textbf{2}; \textbf{3}, \textbf{2}; \textbf{2}_1, \textbf{2})$ \\
        55 & $(\textbf{2}, \textbf{2}, \textbf{2}; \textbf{3}, \textbf{2}; \textbf{2}_2, \textbf{2})$ \\
        56 & $(\textbf{2}, \textbf{2}, \textbf{2}; \textbf{3}, \textbf{2}; \textbf{4}, \textbf{4})$ \\
        57 & $(\textbf{2}, \textbf{3}, \textbf{1}; \textbf{1}, \textbf{2}; \textbf{2}, \textbf{2})$ \\
        58 & $(\textbf{2}, \textbf{3}, \textbf{1}; \textbf{2}, \textbf{1}; \textbf{2}, \textbf{2})$ \\
        59 & $(\textbf{2}, \textbf{3}, \textbf{1}; \textbf{2}, \textbf{3}; \textbf{2}, \textbf{2})$ \\
        60 & $(\textbf{2}, \textbf{3}, \textbf{1}; \textbf{2}, \textbf{3}; \textbf{4}, \textbf{4})$ \\
        61 & $(\textbf{2}, \textbf{3}, \textbf{1}; \textbf{3}, \textbf{2}; \textbf{2}, \textbf{2})$ \\
        62 & $(\textbf{2}, \textbf{3}, \textbf{1}; \textbf{3}, \textbf{2}; \textbf{4}, \textbf{4})$ \\
        \hline
    \end{tabular}
    \begin{tabular}{|c|c|}
    \hline
         & symbol\\\hline
        63 & $(\textbf{2}, \textbf{3}, \textbf{2}; \textbf{1}, \textbf{1}; \textbf{1}, \textbf{1})$ \\
        64 & $(\textbf{2}, \textbf{3}, \textbf{2}; \textbf{1}, \textbf{3}; \textbf{3}_1, \textbf{3})$ \\
        65 & $(\textbf{2}, \textbf{3}, \textbf{2}; \textbf{1}, \textbf{3}; \textbf{3}_2, \textbf{3})$ \\
        66 & $(\textbf{2}, \textbf{3}, \textbf{2}; \textbf{2}, \textbf{2}; \textbf{1}, \textbf{1})$ \\
        67 & $(\textbf{2}, \textbf{3}, \textbf{2}; \textbf{2}, \textbf{2}; \textbf{3}_1, \textbf{3})$ \\
        68 & $(\textbf{2}, \textbf{3}, \textbf{2}; \textbf{2}, \textbf{2}; \textbf{3}_2, \textbf{3})$ \\
        69 & $(\textbf{2}, \textbf{3}, \textbf{2}; \textbf{3}, \textbf{1}; \textbf{3}_1, \textbf{3})$ \\
        70 & $(\textbf{2}, \textbf{3}, \textbf{2}; \textbf{3}, \textbf{1}; \textbf{3}_2, \textbf{3})$ \\
        71 & $(\textbf{2}, \textbf{3}, \textbf{2}; \textbf{3}, \textbf{3}; \textbf{1}, \textbf{1})$ \\
        72 & $(\textbf{2}, \textbf{3}, \textbf{2}; \textbf{3}, \textbf{3}; \textbf{3}_1, \textbf{3})$ \\
        73 & $(\textbf{2}, \textbf{3}, \textbf{2}; \textbf{3}, \textbf{3}; \textbf{3}_2, \textbf{3})$ \\
        74 & $(\textbf{2}, \textbf{3}, \textbf{2}; \textbf{3}, \textbf{3}; \textbf{5}, \textbf{5})$ \\
        75 & $(\textbf{3}, \textbf{1}, \textbf{1}; \textbf{1}, \textbf{3}; \textbf{3}, \textbf{3})$ \\
        76 & $(\textbf{3}, \textbf{1}, \textbf{1}; \textbf{2}, \textbf{2}; \textbf{3}, \textbf{3})$ \\
        77 & $(\textbf{3}, \textbf{1}, \textbf{1}; \textbf{3}, \textbf{1}; \textbf{3}, \textbf{3})$ \\
        78 & $(\textbf{3}, \textbf{1}, \textbf{1}; \textbf{3}, \textbf{3}; \textbf{3}, \textbf{3})$ \\
        79 & $(\textbf{3}, \textbf{1}, \textbf{2}; \textbf{1}, \textbf{2}; \textbf{2}, \textbf{2})$ \\
        80 & $(\textbf{3}, \textbf{1}, \textbf{2}; \textbf{2}, \textbf{1}; \textbf{2}, \textbf{2})$ \\
        81 & $(\textbf{3}, \textbf{1}, \textbf{2}; \textbf{2}, \textbf{3}; \textbf{2}, \textbf{2})$ \\
        82 & $(\textbf{3}, \textbf{1}, \textbf{2}; \textbf{2}, \textbf{3}; \textbf{4}, \textbf{4})$ \\
        83 & $(\textbf{3}, \textbf{1}, \textbf{2}; \textbf{3}, \textbf{2}; \textbf{2}, \textbf{2})$ \\
        84 & $(\textbf{3}, \textbf{1}, \textbf{2}; \textbf{3}, \textbf{2}; \textbf{4}, \textbf{4})$ \\
        85 & $(\textbf{3}, \textbf{2}, \textbf{1}; \textbf{1}, \textbf{2}; \textbf{2}, \textbf{2})$ \\
        86 & $(\textbf{3}, \textbf{2}, \textbf{1}; \textbf{2}, \textbf{1}; \textbf{2}, \textbf{2})$ \\
        87 & $(\textbf{3}, \textbf{2}, \textbf{1}; \textbf{2}, \textbf{3}; \textbf{2}, \textbf{2})$ \\
        88 & $(\textbf{3}, \textbf{2}, \textbf{1}; \textbf{2}, \textbf{3}; \textbf{4}, \textbf{4})$ \\
        89 & $(\textbf{3}, \textbf{2}, \textbf{1}; \textbf{3}, \textbf{2}; \textbf{2}, \textbf{2})$ \\
        90 & $(\textbf{3}, \textbf{2}, \textbf{1}; \textbf{3}, \textbf{2}; \textbf{4}, \textbf{4})$ \\
        91 & $(\textbf{3}, \textbf{2}, \textbf{2}; \textbf{1}, \textbf{1}; \textbf{1}, \textbf{1})$ \\
        92 & $(\textbf{3}, \textbf{2}, \textbf{2}; \textbf{1}, \textbf{3}; \textbf{3}_1, \textbf{3})$ \\
        93 & $(\textbf{3}, \textbf{2}, \textbf{2}; \textbf{1}, \textbf{3}; \textbf{3}_2, \textbf{3})$ \\
        \hline
    \end{tabular}
    \begin{tabular}{|c|c|}
    \hline
         & symbol\\\hline
        94 & $(\textbf{3}, \textbf{2}, \textbf{2}; \textbf{2}, \textbf{2}; \textbf{1}, \textbf{1})$ \\
        95 & $(\textbf{3}, \textbf{2}, \textbf{2}; \textbf{2}, \textbf{2}; \textbf{3}_1, \textbf{3})$ \\
        96 & $(\textbf{3}, \textbf{2}, \textbf{2}; \textbf{2}, \textbf{2}; \textbf{3}_2, \textbf{3})$ \\
        97 & $(\textbf{3}, \textbf{2}, \textbf{2}; \textbf{3}, \textbf{1}; \textbf{3}_1, \textbf{3})$ \\
        98 & $(\textbf{3}, \textbf{2}, \textbf{2}; \textbf{3}, \textbf{1}; \textbf{3}_2, \textbf{3})$ \\
        99 & $(\textbf{3}, \textbf{2}, \textbf{2}; \textbf{3}, \textbf{3}; \textbf{1}, \textbf{1})$ \\
        100 & $(\textbf{3}, \textbf{2}, \textbf{2}; \textbf{3}, \textbf{3}; \textbf{3}_1, \textbf{3})$ \\
        101 & $(\textbf{3}, \textbf{2}, \textbf{2}; \textbf{3}, \textbf{3}; \textbf{3}_2, \textbf{3})$ \\
        102 & $(\textbf{3}, \textbf{2}, \textbf{2}; \textbf{3}, \textbf{3}; \textbf{5}, \textbf{5})$ \\
        103 & $(\textbf{3}, \textbf{3}, \textbf{1}; \textbf{1}, \textbf{1}; \textbf{1}, \textbf{1})$ \\
        104 & $(\textbf{3}, \textbf{3}, \textbf{1}; \textbf{1}, \textbf{3}; \textbf{3}, \textbf{3})$ \\
        105 & $(\textbf{3}, \textbf{3}, \textbf{1}; \textbf{2}, \textbf{2}; \textbf{1}, \textbf{1})$ \\
        106 & $(\textbf{3}, \textbf{3}, \textbf{1}; \textbf{2}, \textbf{2}; \textbf{3}, \textbf{3})$ \\
        107 & $(\textbf{3}, \textbf{3}, \textbf{1}; \textbf{3}, \textbf{1}; \textbf{3}, \textbf{3})$ \\
        108 & $(\textbf{3}, \textbf{3}, \textbf{1}; \textbf{3}, \textbf{3}; \textbf{1}, \textbf{1})$ \\
        109 & $(\textbf{3}, \textbf{3}, \textbf{1}; \textbf{3}, \textbf{3}; \textbf{3}, \textbf{3})$ \\
        110 & $(\textbf{3}, \textbf{3}, \textbf{1}; \textbf{3}, \textbf{3}; \textbf{5}, \textbf{5})$ \\
        111 & $(\textbf{3}, \textbf{3}, \textbf{2}; \textbf{1}, \textbf{2}; \textbf{2}_1, \textbf{2})$ \\
        112 & $(\textbf{3}, \textbf{3}, \textbf{2}; \textbf{1}, \textbf{2}; \textbf{2}_2, \textbf{2})$ \\
        113 & $(\textbf{3}, \textbf{3}, \textbf{2}; \textbf{2}, \textbf{1}; \textbf{2}_1, \textbf{2})$ \\
        114 & $(\textbf{3}, \textbf{3}, \textbf{2}; \textbf{2}, \textbf{1}; \textbf{2}_2, \textbf{2})$ \\
        115 & $(\textbf{3}, \textbf{3}, \textbf{2}; \textbf{2}, \textbf{3}; \textbf{2}_1, \textbf{2})$ \\
        116 & $(\textbf{3}, \textbf{3}, \textbf{2}; \textbf{2}, \textbf{3}; \textbf{2}_2, \textbf{2})$ \\
        117 & $(\textbf{3}, \textbf{3}, \textbf{2}; \textbf{2}, \textbf{3}; \textbf{4}_1, \textbf{4})$ \\
        118 & $(\textbf{3}, \textbf{3}, \textbf{2}; \textbf{2}, \textbf{3}; \textbf{4}_2, \textbf{4})$ \\
        119 & $(\textbf{3}, \textbf{3}, \textbf{2}; \textbf{3}, \textbf{2}; \textbf{2}_1, \textbf{2})$ \\
        120 & $(\textbf{3}, \textbf{3}, \textbf{2}; \textbf{3}, \textbf{2}; \textbf{2}_2, \textbf{2})$ \\
        121 & $(\textbf{3}, \textbf{3}, \textbf{2}; \textbf{3}, \textbf{2}; \textbf{4}_1, \textbf{4})$ \\
        122 & $(\textbf{3}, \textbf{3}, \textbf{2}; \textbf{3}, \textbf{2}; \textbf{4}_2, \textbf{4})$ \\
        &\\
        &\\
        \hline
    \end{tabular}
    }
    \caption{5-representation indices for SU(2). The subscript denotes the multiplicity index (if any).}
    \label{tab:su2_poly}
\end{table}
\begin{table}
    \centering
    {\scriptsize
    \begin{tabular}{|c|c|}
    \hline
        index & symbol\\\hline
        1 & $(\textbf{1}, \textbf{1}, \textbf{1}; \textbf{1}, \textbf{1}; \textbf{1}, \textbf{1})$ \\
        2 & $(\textbf{1}, \textbf{1}, \textbf{1}; \textbf{3}, \bar{\textbf{3}}; \textbf{1}, \textbf{1})$ \\
        3 & $(\textbf{1}, \textbf{1}, \textbf{1}; \bar{\textbf{3}}, \textbf{3}; \textbf{1}, \textbf{1})$ \\
        4 & $(\textbf{1}, \textbf{1}, \textbf{3}; \textbf{1}, \textbf{3}; \textbf{3}, \textbf{3})$ \\
        5 & $(\textbf{1}, \textbf{1}, \textbf{3}; \textbf{3}, \textbf{1}; \textbf{3}, \textbf{3})$ \\
        6 & $(\textbf{1}, \textbf{1}, \textbf{3}; \bar{\textbf{3}}, \bar{\textbf{3}}; \textbf{3}, \textbf{3})$ \\
        7 & $(\textbf{1}, \textbf{1}, \bar{\textbf{3}}; \textbf{1}, \bar{\textbf{3}}; \bar{\textbf{3}}, \bar{\textbf{3}})$ \\
        8 & $(\textbf{1}, \textbf{1}, \bar{\textbf{3}}; \textbf{3}, \textbf{3}; \bar{\textbf{3}}, \bar{\textbf{3}})$ \\
        9 & $(\textbf{1}, \textbf{1}, \bar{\textbf{3}}; \bar{\textbf{3}}, \textbf{1}; \bar{\textbf{3}}, \bar{\textbf{3}})$ \\
        10 & $(\textbf{1}, \textbf{3}, \textbf{1}; \textbf{1}, \textbf{3}; \textbf{3}, \textbf{3})$ \\
        11 & $(\textbf{1}, \textbf{3}, \textbf{1}; \textbf{3}, \textbf{1}; \textbf{3}, \textbf{3})$ \\
        12 & $(\textbf{1}, \textbf{3}, \textbf{1}; \bar{\textbf{3}}, \bar{\textbf{3}}; \textbf{3}, \textbf{3})$ \\
        13 & $(\textbf{1}, \textbf{3}, \textbf{3}; \textbf{1}, \bar{\textbf{3}}; \bar{\textbf{3}}, \bar{\textbf{3}})$ \\
        14 & $(\textbf{1}, \textbf{3}, \textbf{3}; \textbf{3}, \textbf{3}; \bar{\textbf{3}}, \bar{\textbf{3}})$ \\
        15 & $(\textbf{1}, \textbf{3}, \textbf{3}; \textbf{3}, \textbf{3}; \textbf{6}, \textbf{6})$ \\
        16 & $(\textbf{1}, \textbf{3}, \textbf{3}; \bar{\textbf{3}}, \textbf{1}; \bar{\textbf{3}}, \bar{\textbf{3}})$ \\
        17 & $(\textbf{1}, \textbf{3}, \bar{\textbf{3}}; \textbf{1}, \textbf{1}; \textbf{1}, \textbf{1})$ \\
        18 & $(\textbf{1}, \textbf{3}, \bar{\textbf{3}}; \textbf{3}, \bar{\textbf{3}}; \textbf{1}, \textbf{1})$ \\
        19 & $(\textbf{1}, \textbf{3}, \bar{\textbf{3}}; \textbf{3}, \bar{\textbf{3}}; \textbf{8}, \textbf{8})$ \\
        20 & $(\textbf{1}, \textbf{3}, \bar{\textbf{3}}; \bar{\textbf{3}}, \textbf{3}; \textbf{1}, \textbf{1})$ \\
        21 & $(\textbf{1}, \textbf{3}, \bar{\textbf{3}}; \bar{\textbf{3}}, \textbf{3}; \textbf{8}, \textbf{8})$ \\
        22 & $(\textbf{1}, \bar{\textbf{3}}, \textbf{1}; \textbf{1}, \bar{\textbf{3}}; \bar{\textbf{3}}, \bar{\textbf{3}})$ \\
        23 & $(\textbf{1}, \bar{\textbf{3}}, \textbf{1}; \textbf{3}, \textbf{3}; \bar{\textbf{3}}, \bar{\textbf{3}})$ \\
        24 & $(\textbf{1}, \bar{\textbf{3}}, \textbf{1}; \bar{\textbf{3}}, \textbf{1}; \bar{\textbf{3}}, \bar{\textbf{3}})$ \\
        25 & $(\textbf{1}, \bar{\textbf{3}}, \textbf{3}; \textbf{1}, \textbf{1}; \textbf{1}, \textbf{1})$ \\
        26 & $(\textbf{1}, \bar{\textbf{3}}, \textbf{3}; \textbf{3}, \bar{\textbf{3}}; \textbf{1}, \textbf{1})$ \\
        27 & $(\textbf{1}, \bar{\textbf{3}}, \textbf{3}; \textbf{3}, \bar{\textbf{3}}; \textbf{8}, \textbf{8})$ \\
        28 & $(\textbf{1}, \bar{\textbf{3}}, \textbf{3}; \bar{\textbf{3}}, \textbf{3}; \textbf{1}, \textbf{1})$ \\
        29 & $(\textbf{1}, \bar{\textbf{3}}, \textbf{3}; \bar{\textbf{3}}, \textbf{3}; \textbf{8}, \textbf{8})$ \\
        30 & $(\textbf{1}, \bar{\textbf{3}}, \bar{\textbf{3}}; \textbf{1}, \textbf{3}; \textbf{3}, \textbf{3})$ \\
        31 & $(\textbf{1}, \bar{\textbf{3}}, \bar{\textbf{3}}; \textbf{3}, \textbf{1}; \textbf{3}, \textbf{3})$ \\
        32 & $(\textbf{1}, \bar{\textbf{3}}, \bar{\textbf{3}}; \bar{\textbf{3}}, \bar{\textbf{3}}; \textbf{3}, \textbf{3})$ \\
        33 & $(\textbf{1}, \bar{\textbf{3}}, \bar{\textbf{3}}; \bar{\textbf{3}}, \bar{\textbf{3}}; \bar{\textbf{6}}, \bar{\textbf{6}})$ \\
        \hline
    \end{tabular}
    \begin{tabular}{|c|c|}
    \hline
         & symbol\\\hline
        34 & $(\textbf{3}, \textbf{1}, \textbf{1}; \textbf{1}, \textbf{3}; \textbf{3}, \textbf{3})$ \\
        35 & $(\textbf{3}, \textbf{1}, \textbf{1}; \textbf{3}, \textbf{1}; \textbf{3}, \textbf{3})$ \\
        36 & $(\textbf{3}, \textbf{1}, \textbf{1}; \bar{\textbf{3}}, \bar{\textbf{3}}; \textbf{3}, \textbf{3})$ \\
        37 & $(\textbf{3}, \textbf{1}, \textbf{3}; \textbf{1}, \bar{\textbf{3}}; \bar{\textbf{3}}, \bar{\textbf{3}})$ \\
        38 & $(\textbf{3}, \textbf{1}, \textbf{3}; \textbf{3}, \textbf{3}; \bar{\textbf{3}}, \bar{\textbf{3}})$ \\
        39 & $(\textbf{3}, \textbf{1}, \textbf{3}; \textbf{3}, \textbf{3}; \textbf{6}, \textbf{6})$ \\
        40 & $(\textbf{3}, \textbf{1}, \textbf{3}; \bar{\textbf{3}}, \textbf{1}; \bar{\textbf{3}}, \bar{\textbf{3}})$ \\
        41 & $(\textbf{3}, \textbf{1}, \bar{\textbf{3}}; \textbf{1}, \textbf{1}; \textbf{1}, \textbf{1})$ \\
        42 & $(\textbf{3}, \textbf{1}, \bar{\textbf{3}}; \textbf{3}, \bar{\textbf{3}}; \textbf{1}, \textbf{1})$ \\
        43 & $(\textbf{3}, \textbf{1}, \bar{\textbf{3}}; \textbf{3}, \bar{\textbf{3}}; \textbf{8}, \textbf{8})$ \\
        44 & $(\textbf{3}, \textbf{1}, \bar{\textbf{3}}; \bar{\textbf{3}}, \textbf{3}; \textbf{1}, \textbf{1})$ \\
        45 & $(\textbf{3}, \textbf{1}, \bar{\textbf{3}}; \bar{\textbf{3}}, \textbf{3}; \textbf{8}, \textbf{8})$ \\
        46 & $(\textbf{3}, \textbf{3}, \textbf{1}; \textbf{1}, \bar{\textbf{3}}; \bar{\textbf{3}}, \bar{\textbf{3}})$ \\
        47 & $(\textbf{3}, \textbf{3}, \textbf{1}; \textbf{3}, \textbf{3}; \bar{\textbf{3}}, \bar{\textbf{3}})$ \\
        48 & $(\textbf{3}, \textbf{3}, \textbf{1}; \textbf{3}, \textbf{3}; \textbf{6}, \textbf{6})$ \\
        49 & $(\textbf{3}, \textbf{3}, \textbf{1}; \bar{\textbf{3}}, \textbf{1}; \bar{\textbf{3}}, \bar{\textbf{3}})$ \\
        50 & $(\textbf{3}, \textbf{3}, \textbf{3}; \textbf{1}, \textbf{1}; \textbf{1}, \textbf{1})$ \\
        51 & $(\textbf{3}, \textbf{3}, \textbf{3}; \textbf{3}, \bar{\textbf{3}}; \textbf{1}, \textbf{1})$ \\
        52 & $(\textbf{3}, \textbf{3}, \textbf{3}; \textbf{3}, \bar{\textbf{3}}; \textbf{8}_1, \textbf{8})$ \\
        53 & $(\textbf{3}, \textbf{3}, \textbf{3}; \textbf{3}, \bar{\textbf{3}}; \textbf{8}_2, \textbf{8})$ \\
        54 & $(\textbf{3}, \textbf{3}, \textbf{3}; \bar{\textbf{3}}, \textbf{3}; \textbf{1}, \textbf{1})$ \\
        55 & $(\textbf{3}, \textbf{3}, \textbf{3}; \bar{\textbf{3}}, \textbf{3}; \textbf{8}_1, \textbf{8})$ \\
        56 & $(\textbf{3}, \textbf{3}, \textbf{3}; \bar{\textbf{3}}, \textbf{3}; \textbf{8}_2, \textbf{8})$ \\
        57 & $(\textbf{3}, \textbf{3}, \bar{\textbf{3}}; \textbf{1}, \textbf{3}; \textbf{3}_1, \textbf{3})$ \\
        58 & $(\textbf{3}, \textbf{3}, \bar{\textbf{3}}; \textbf{1}, \textbf{3}; \textbf{3}_2, \textbf{3})$ \\
        59 & $(\textbf{3}, \textbf{3}, \bar{\textbf{3}}; \textbf{3}, \textbf{1}; \textbf{3}_1, \textbf{3})$ \\
        60 & $(\textbf{3}, \textbf{3}, \bar{\textbf{3}}; \textbf{3}, \textbf{1}; \textbf{3}_2, \textbf{3})$ \\
        61 & $(\textbf{3}, \textbf{3}, \bar{\textbf{3}}; \bar{\textbf{3}}, \bar{\textbf{3}}; \textbf{3}_1, \textbf{3})$ \\
        62 & $(\textbf{3}, \textbf{3}, \bar{\textbf{3}}; \bar{\textbf{3}}, \bar{\textbf{3}}; \textbf{3}_2, \textbf{3})$ \\
        63 & $(\textbf{3}, \textbf{3}, \bar{\textbf{3}}; \bar{\textbf{3}}, \bar{\textbf{3}}; \bar{\textbf{6}}, \bar{\textbf{6}})$ \\
        64 & $(\textbf{3}, \bar{\textbf{3}}, \textbf{1}; \textbf{1}, \textbf{1}; \textbf{1}, \textbf{1})$ \\
        65 & $(\textbf{3}, \bar{\textbf{3}}, \textbf{1}; \textbf{3}, \bar{\textbf{3}}; \textbf{1}, \textbf{1})$ \\
        66 & $(\textbf{3}, \bar{\textbf{3}}, \textbf{1}; \textbf{3}, \bar{\textbf{3}}; \textbf{8}, \textbf{8})$ \\
        \hline
    \end{tabular}
    \begin{tabular}{|c|c|}
    \hline
         & symbol\\\hline
        67 & $(\textbf{3}, \bar{\textbf{3}}, \textbf{1}; \bar{\textbf{3}}, \textbf{3}; \textbf{1}, \textbf{1})$ \\
        68 & $(\textbf{3}, \bar{\textbf{3}}, \textbf{1}; \bar{\textbf{3}}, \textbf{3}; \textbf{8}, \textbf{8})$ \\
        69 & $(\textbf{3}, \bar{\textbf{3}}, \textbf{3}; \textbf{1}, \textbf{3}; \textbf{3}_1, \textbf{3})$ \\
        70 & $(\textbf{3}, \bar{\textbf{3}}, \textbf{3}; \textbf{1}, \textbf{3}; \textbf{3}_2, \textbf{3})$ \\
        71 & $(\textbf{3}, \bar{\textbf{3}}, \textbf{3}; \textbf{3}, \textbf{1}; \textbf{3}_1, \textbf{3})$ \\
        72 & $(\textbf{3}, \bar{\textbf{3}}, \textbf{3}; \textbf{3}, \textbf{1}; \textbf{3}_2, \textbf{3})$ \\
        73 & $(\textbf{3}, \bar{\textbf{3}}, \textbf{3}; \bar{\textbf{3}}, \bar{\textbf{3}}; \textbf{3}_1, \textbf{3})$ \\
        74 & $(\textbf{3}, \bar{\textbf{3}}, \textbf{3}; \bar{\textbf{3}}, \bar{\textbf{3}}; \textbf{3}_2, \textbf{3})$ \\
        75 & $(\textbf{3}, \bar{\textbf{3}}, \textbf{3}; \bar{\textbf{3}}, \bar{\textbf{3}}; \bar{\textbf{6}}, \bar{\textbf{6}})$ \\
        76 & $(\textbf{3}, \bar{\textbf{3}}, \bar{\textbf{3}}; \textbf{1}, \bar{\textbf{3}}; \bar{\textbf{3}}_1, \bar{\textbf{3}})$ \\
        77 & $(\textbf{3}, \bar{\textbf{3}}, \bar{\textbf{3}}; \textbf{1}, \bar{\textbf{3}}; \bar{\textbf{3}}_2, \bar{\textbf{3}})$ \\
        78 & $(\textbf{3}, \bar{\textbf{3}}, \bar{\textbf{3}}; \textbf{3}, \textbf{3}; \bar{\textbf{3}}_1, \bar{\textbf{3}})$ \\
        79 & $(\textbf{3}, \bar{\textbf{3}}, \bar{\textbf{3}}; \textbf{3}, \textbf{3}; \bar{\textbf{3}}_2, \bar{\textbf{3}})$ \\
        80 & $(\textbf{3}, \bar{\textbf{3}}, \bar{\textbf{3}}; \textbf{3}, \textbf{3}; \textbf{6}, \textbf{6})$ \\
        81 & $(\textbf{3}, \bar{\textbf{3}}, \bar{\textbf{3}}; \bar{\textbf{3}}, \textbf{1}; \bar{\textbf{3}}_1, \bar{\textbf{3}})$ \\
        82 & $(\textbf{3}, \bar{\textbf{3}}, \bar{\textbf{3}}; \bar{\textbf{3}}, \textbf{1}; \bar{\textbf{3}}_2, \bar{\textbf{3}})$ \\
        83 & $(\bar{\textbf{3}}, \textbf{1}, \textbf{1}; \textbf{1}, \bar{\textbf{3}}; \bar{\textbf{3}}, \bar{\textbf{3}})$ \\
        84 & $(\bar{\textbf{3}}, \textbf{1}, \textbf{1}; \textbf{3}, \textbf{3}; \bar{\textbf{3}}, \bar{\textbf{3}})$ \\
        85 & $(\bar{\textbf{3}}, \textbf{1}, \textbf{1}; \bar{\textbf{3}}, \textbf{1}; \bar{\textbf{3}}, \bar{\textbf{3}})$ \\
        86 & $(\bar{\textbf{3}}, \textbf{1}, \textbf{3}; \textbf{1}, \textbf{1}; \textbf{1}, \textbf{1})$ \\
        87 & $(\bar{\textbf{3}}, \textbf{1}, \textbf{3}; \textbf{3}, \bar{\textbf{3}}; \textbf{1}, \textbf{1})$ \\
        88 & $(\bar{\textbf{3}}, \textbf{1}, \textbf{3}; \textbf{3}, \bar{\textbf{3}}; \textbf{8}, \textbf{8})$ \\
        89 & $(\bar{\textbf{3}}, \textbf{1}, \textbf{3}; \bar{\textbf{3}}, \textbf{3}; \textbf{1}, \textbf{1})$ \\
        90 & $(\bar{\textbf{3}}, \textbf{1}, \textbf{3}; \bar{\textbf{3}}, \textbf{3}; \textbf{8}, \textbf{8})$ \\
        91 & $(\bar{\textbf{3}}, \textbf{1}, \bar{\textbf{3}}; \textbf{1}, \textbf{3}; \textbf{3}, \textbf{3})$ \\
        92 & $(\bar{\textbf{3}}, \textbf{1}, \bar{\textbf{3}}; \textbf{3}, \textbf{1}; \textbf{3}, \textbf{3})$ \\
        93 & $(\bar{\textbf{3}}, \textbf{1}, \bar{\textbf{3}}; \bar{\textbf{3}}, \bar{\textbf{3}}; \textbf{3}, \textbf{3})$ \\
        94 & $(\bar{\textbf{3}}, \textbf{1}, \bar{\textbf{3}}; \bar{\textbf{3}}, \bar{\textbf{3}}; \bar{\textbf{6}}, \bar{\textbf{6}})$ \\
        95 & $(\bar{\textbf{3}}, \textbf{3}, \textbf{1}; \textbf{1}, \textbf{1}; \textbf{1}, \textbf{1})$ \\
        96 & $(\bar{\textbf{3}}, \textbf{3}, \textbf{1}; \textbf{3}, \bar{\textbf{3}}; \textbf{1}, \textbf{1})$ \\
        97 & $(\bar{\textbf{3}}, \textbf{3}, \textbf{1}; \textbf{3}, \bar{\textbf{3}}; \textbf{8}, \textbf{8})$ \\
        98 & $(\bar{\textbf{3}}, \textbf{3}, \textbf{1}; \bar{\textbf{3}}, \textbf{3}; \textbf{1}, \textbf{1})$ \\
        99 & $(\bar{\textbf{3}}, \textbf{3}, \textbf{1}; \bar{\textbf{3}}, \textbf{3}; \textbf{8}, \textbf{8})$ \\
        \hline
    \end{tabular}
    \begin{tabular}{|c|c|}
    \hline
         & symbol\\\hline
        100 & $(\bar{\textbf{3}}, \textbf{3}, \textbf{3}; \textbf{1}, \textbf{3}; \textbf{3}_1, \textbf{3})$ \\
        101 & $(\bar{\textbf{3}}, \textbf{3}, \textbf{3}; \textbf{1}, \textbf{3}; \textbf{3}_2, \textbf{3})$ \\
        102 & $(\bar{\textbf{3}}, \textbf{3}, \textbf{3}; \textbf{3}, \textbf{1}; \textbf{3}_1, \textbf{3})$ \\
        103 & $(\bar{\textbf{3}}, \textbf{3}, \textbf{3}; \textbf{3}, \textbf{1}; \textbf{3}_2, \textbf{3})$ \\
        104 & $(\bar{\textbf{3}}, \textbf{3}, \textbf{3}; \bar{\textbf{3}}, \bar{\textbf{3}}; \textbf{3}_1, \textbf{3})$ \\
        105 & $(\bar{\textbf{3}}, \textbf{3}, \textbf{3}; \bar{\textbf{3}}, \bar{\textbf{3}}; \textbf{3}_2, \textbf{3})$ \\
        106 & $(\bar{\textbf{3}}, \textbf{3}, \textbf{3}; \bar{\textbf{3}}, \bar{\textbf{3}}; \bar{\textbf{6}}, \bar{\textbf{6}})$ \\
        107 & $(\bar{\textbf{3}}, \textbf{3}, \bar{\textbf{3}}; \textbf{1}, \bar{\textbf{3}}; \bar{\textbf{3}}_1, \bar{\textbf{3}})$ \\
        108 & $(\bar{\textbf{3}}, \textbf{3}, \bar{\textbf{3}}; \textbf{1}, \bar{\textbf{3}}; \bar{\textbf{3}}_2, \bar{\textbf{3}})$ \\
        109 & $(\bar{\textbf{3}}, \textbf{3}, \bar{\textbf{3}}; \textbf{3}, \textbf{3}; \bar{\textbf{3}}_1, \bar{\textbf{3}})$ \\
        110 & $(\bar{\textbf{3}}, \textbf{3}, \bar{\textbf{3}}; \textbf{3}, \textbf{3}; \bar{\textbf{3}}_2, \bar{\textbf{3}})$ \\
        111 & $(\bar{\textbf{3}}, \textbf{3}, \bar{\textbf{3}}; \textbf{3}, \textbf{3}; \textbf{6}, \textbf{6})$ \\
        112 & $(\bar{\textbf{3}}, \textbf{3}, \bar{\textbf{3}}; \bar{\textbf{3}}, \textbf{1}; \bar{\textbf{3}}_1, \bar{\textbf{3}})$ \\
        113 & $(\bar{\textbf{3}}, \textbf{3}, \bar{\textbf{3}}; \bar{\textbf{3}}, \textbf{1}; \bar{\textbf{3}}_2, \bar{\textbf{3}})$ \\
        114 & $(\bar{\textbf{3}}, \bar{\textbf{3}}, \textbf{1}; \textbf{1}, \textbf{3}; \textbf{3}, \textbf{3})$ \\
        115 & $(\bar{\textbf{3}}, \bar{\textbf{3}}, \textbf{1}; \textbf{3}, \textbf{1}; \textbf{3}, \textbf{3})$ \\
        116 & $(\bar{\textbf{3}}, \bar{\textbf{3}}, \textbf{1}; \bar{\textbf{3}}, \bar{\textbf{3}}; \textbf{3}, \textbf{3})$ \\
        117 & $(\bar{\textbf{3}}, \bar{\textbf{3}}, \textbf{1}; \bar{\textbf{3}}, \bar{\textbf{3}}; \bar{\textbf{6}}, \bar{\textbf{6}})$ \\
        118 & $(\bar{\textbf{3}}, \bar{\textbf{3}}, \textbf{3}; \textbf{1}, \bar{\textbf{3}}; \bar{\textbf{3}}_1, \bar{\textbf{3}})$ \\
        119 & $(\bar{\textbf{3}}, \bar{\textbf{3}}, \textbf{3}; \textbf{1}, \bar{\textbf{3}}; \bar{\textbf{3}}_2, \bar{\textbf{3}})$ \\
        120 & $(\bar{\textbf{3}}, \bar{\textbf{3}}, \textbf{3}; \textbf{3}, \textbf{3}; \bar{\textbf{3}}_1, \bar{\textbf{3}})$ \\
        121 & $(\bar{\textbf{3}}, \bar{\textbf{3}}, \textbf{3}; \textbf{3}, \textbf{3}; \bar{\textbf{3}}_2, \bar{\textbf{3}})$ \\
        122 & $(\bar{\textbf{3}}, \bar{\textbf{3}}, \textbf{3}; \textbf{3}, \textbf{3}; \textbf{6}, \textbf{6})$ \\
        123 & $(\bar{\textbf{3}}, \bar{\textbf{3}}, \textbf{3}; \bar{\textbf{3}}, \textbf{1}; \bar{\textbf{3}}_1, \bar{\textbf{3}})$ \\
        124 & $(\bar{\textbf{3}}, \bar{\textbf{3}}, \textbf{3}; \bar{\textbf{3}}, \textbf{1}; \bar{\textbf{3}}_2, \bar{\textbf{3}})$ \\
        125 & $(\bar{\textbf{3}}, \bar{\textbf{3}}, \bar{\textbf{3}}; \textbf{1}, \textbf{1}; \textbf{1}, \textbf{1})$ \\
        126 & $(\bar{\textbf{3}}, \bar{\textbf{3}}, \bar{\textbf{3}}; \textbf{3}, \bar{\textbf{3}}; \textbf{1}, \textbf{1})$ \\
        127 & $(\bar{\textbf{3}}, \bar{\textbf{3}}, \bar{\textbf{3}}; \textbf{3}, \bar{\textbf{3}}; \textbf{8}_1, \textbf{8})$ \\
        128 & $(\bar{\textbf{3}}, \bar{\textbf{3}}, \bar{\textbf{3}}; \textbf{3}, \bar{\textbf{3}}; \textbf{8}_2, \textbf{8})$ \\
        129 & $(\bar{\textbf{3}}, \bar{\textbf{3}}, \bar{\textbf{3}}; \bar{\textbf{3}}, \textbf{3}; \textbf{1}, \textbf{1})$ \\
        130 & $(\bar{\textbf{3}}, \bar{\textbf{3}}, \bar{\textbf{3}}; \bar{\textbf{3}}, \textbf{3}; \textbf{8}_1, \textbf{8})$ \\
        131 & $(\bar{\textbf{3}}, \bar{\textbf{3}}, \bar{\textbf{3}}; \bar{\textbf{3}}, \textbf{3}; \textbf{8}_2, \textbf{8})$ \\
        &\\
        \hline
    \end{tabular}
    }
    \caption{5-representation indices for SU(3). The subscript denotes the multiplicity index (if any).}
    \label{tab:su3_poly}
\end{table}
\bibliographystyle{JHEP}
\bibliography{ref}

\end{document}